\documentclass[12pt]{article}

\usepackage[utf8]{inputenc}
\usepackage{amsmath,amssymb,amsthm,fullpage}
\usepackage{graphicx}
\usepackage{subcaption}
\usepackage{algorithm}
\usepackage{algorithmic}
\usepackage{enumitem}
\usepackage{hyperref}
\usepackage{authblk}

\title{Persistence and material coherence\\ of a mesoscale ocean eddy}
\author{Michael C. Denes}
\author{Gary Froyland}
\author{Shane R. Keating}
\affil{School of Mathematics and Statistics\\ University of New South Wales\\ Sydney NSW 2052, Australia}
\date{\today}

\begin{document}
\maketitle

\begin{abstract}
Ocean eddies play an important role in the transport and mixing processes of the ocean due to their ability to transport material, heat, salt, and other tracers across large distances. They exhibit at least two timescales; an Eulerian lifetime associated with persistent identifiable signatures in gridded fields such as vorticity or sea-surface height, and multiple Lagrangian or material coherence timescales that are typically much shorter. 
We propose a method to study the multi-timescale material transport, leakage, and entrainment by eddies with their surroundings by constructing sequences of finite-time coherent sets, computed as superlevel sets of dominant eigenfunctions of dynamic Laplace operators.
The dominant eigenvalues of dynamic Laplace operators defined on time intervals of varying length allows us to identify a maximal coherence timescale that minimizes the rate of mass loss over a domain, per unit flow time. We apply the method to examine the persistence and material coherence of an Agulhas ring, an ocean eddy in the South Atlantic ocean, using particle trajectories derived from a $0.1^\circ$ global numerical ocean simulation. 
Using a sequence of sliding windows, the method is able to identify and track a persistent eddy feature for a time much longer than the maximal coherence timescale, and with considerably larger material transport than the corresponding eddy feature identified from purely Eulerian information. 
Furthermore, the median residence times of fluid in the identified feature far exceed the timescale over which fully material motion is guaranteed.
Through residence time calculations, we find that this particular eddy does not exhibit a long-lived coherent inner core and that the bulk of material transport is performed by the quasi-coherent outer ring of the eddy.
\end{abstract}

\section{Introduction}\label{sec:intro}
Fluid structures that remain materially coherent for longer than common dynamical timescales (i.e.\ many Lyapunov times or eddy turnover times) are an effective framework for studying the transport of material by advection \cite{Beron-Vera2013, Liu2019}. Methods for identifying Lagrangian coherent structures (LCSs) have found a wide range of applications in geophysical and astrophysical flows (see \cite{Allshouse2015, Hadjighasem2017} and references therein) and have proven to be a particularly useful tool in studying mesoscale ocean eddies.
The latter are rotating vortices of fluid, strongly constrained by stratification and planetary rotation, with typical horizontal lengthscales of tens to hundreds of kilometers and lifetimes of weeks to months \cite{Chelton2007, Chelton2011}. Mesoscale ocean eddies are found ubiquitously throughout the ocean, and can transport mass, heat, salt, and biogeochemical tracers over large distances \cite{Robinson1983,CheltonGaubeSchlax2011}.
As such, eddies play a key role in the maintenance of the global overturning circulation, large-scale water mass distributions, and ocean biology \cite{Robinson1983, Weijer2002, Biastoch2008a}.

In the traditional Eulerian perspective, eddy-like features are identified from signatures in a fixed frame, such as the vorticity field, sea surface height anomaly (a proxy for surface pressure) or the Okubo-Weiss parameter, which quantifies the local balance between rotation (vorticity) and deformation (strain) \cite{Okubo1970, Weiss1991}. These signatures can propagate and persist for up to several years, and typically define the \emph{Eulerian lifetime} of an eddy. Automated tracking algorithms utilizing the Okubo-Weiss parameter \cite{Isern-Fontanet2003, Isern-Fontanet2006, Chelton2007} or sea-surface height anomaly \cite{Fang2003, Chaigneau2005, Fu2006, Chelton2011} have been used extensively to develop censuses of ocean eddies, their lifetimes, pathways, and characteristics \cite{Chelton2011}. However, as is well-known, the boundaries of these eddy-like features are not materially coherent, \cite{Beron-Vera2013, Haller2013, Liu2019, AndradeCanto2020}, and fluid typically leaks across the identified boundary on timescales much shorter than the Eulerian lifetime of the eddy.

Given the inherent Lagrangian nature of ocean eddies and the inability of Eulerian methods to capture material boundaries, objective LCS-detection methods are natural candidates for tracking eddies and quantifying transport \cite{Beron-Vera2008,Froyland2012, Beron-Vera2013, Froyland2015c, AndradeCanto2020}. In the Lagrangian perspective, eddy-like features are identified by either closed material boundaries that minimize mixing with the surrounding fluid, or regions of the fluid that remain coherent and retain mass as they are advected by the flow.
A wide range of Lagrangian approaches to coherent structure detection have been developed in recent years (see \cite{Allshouse2015, Hadjighasem2017} for a summary) and a number of recent studies have applied these to idealized and real-world examples \cite{Beron-Vera2008, Froyland2010, Beron-Vera2013, Haller2013, Froyland2015a, Froyland2015c, Wang2015, BeronVera2015, Haller2016,Wang2016,Abernathey2018,Froyland2018,Beron-Vera2018,Zhang2019,Sinha2019,Liu2019,AndradeCanto2020}.

Lagrangian analysis of simulated and observed velocity data suggest that ocean eddies consist of an inner core --- which, in some instances, remains coherent for the Eulerian lifetime of the eddy --- and an outer ring containing a mixture of trapped fluid and fluid that has been entrained from the surrounding water as the eddy propagates.
In idealized numerical simulations, \cite{Early2011} found that quasi-stable, isolated ocean eddies can transport fluid within the eddy core, defined as the region interior to the zero contour of relative vorticity, while a surrounding ring with opposite signed relative vorticity entrains and sheds fluid parcels throughout the eddy's lifetime. Using satellite-derived geostrophic velocities, \cite{Wang2015} estimated transport by coherent eddy cores in the South Atlantic defined by coherent material belts \cite{Haller2013} and found that cross-ocean-basin transport by eddy cores was two orders of magnitude smaller than prior estimates based on Eulerian eddy tracking. These results were consistent with the findings of \cite{Abernathey2018}, who performed a similar analysis in the Eastern Pacific using the Lagrangian averaged vorticity deviation (LAVD) \cite{Haller2016}. These authors found that coherent eddy cores contribute less than 1\% of the net transport in the Eastern Pacific, suggesting that most transport by ocean eddies is due to motion outside the eddy cores. 

These estimates place a lower bound on material transport by ocean eddies as they quantify transport due to the coherent core only \cite{Abernathey2018}. An important additional component of transport by ocean eddies is due to entrainment and transport of fluid outside the inner core. We refer to this quasi-coherent region surrounding the core as the \emph{outer ring} (not to be confused with Agulhas rings, Gulf Stream rings, etc., which have a specific oceanographic interpretation). Few studies have attempted to objectively estimate the fluid transport by the outer ring. \cite{Early2011} found that the outer ring of a simulated ocean eddy consisted of a mixture of trapped fluid and fluid entrained from surrounding waters that was transported with the eddy for over 1000 km. \cite{Froyland2015c} studied the decay of an eddy in the South Atlantic ocean using finite-time coherent sets (regions of phase space which experience minimal leakage) computed from transfer operators \cite{Froyland2010} in a windowing scheme that allows for an object to shrink or grow over time. This approach accounts for both the inner core and outer ring as one time-varying object. They found that approximately 15\% of the initial water mass leaked out of the combined eddy core and outer ring by the time it reached $25^\circ$W in the Western Atlantic ocean. 

\begin{figure}[h]
\centering
\includegraphics[width=1\linewidth]{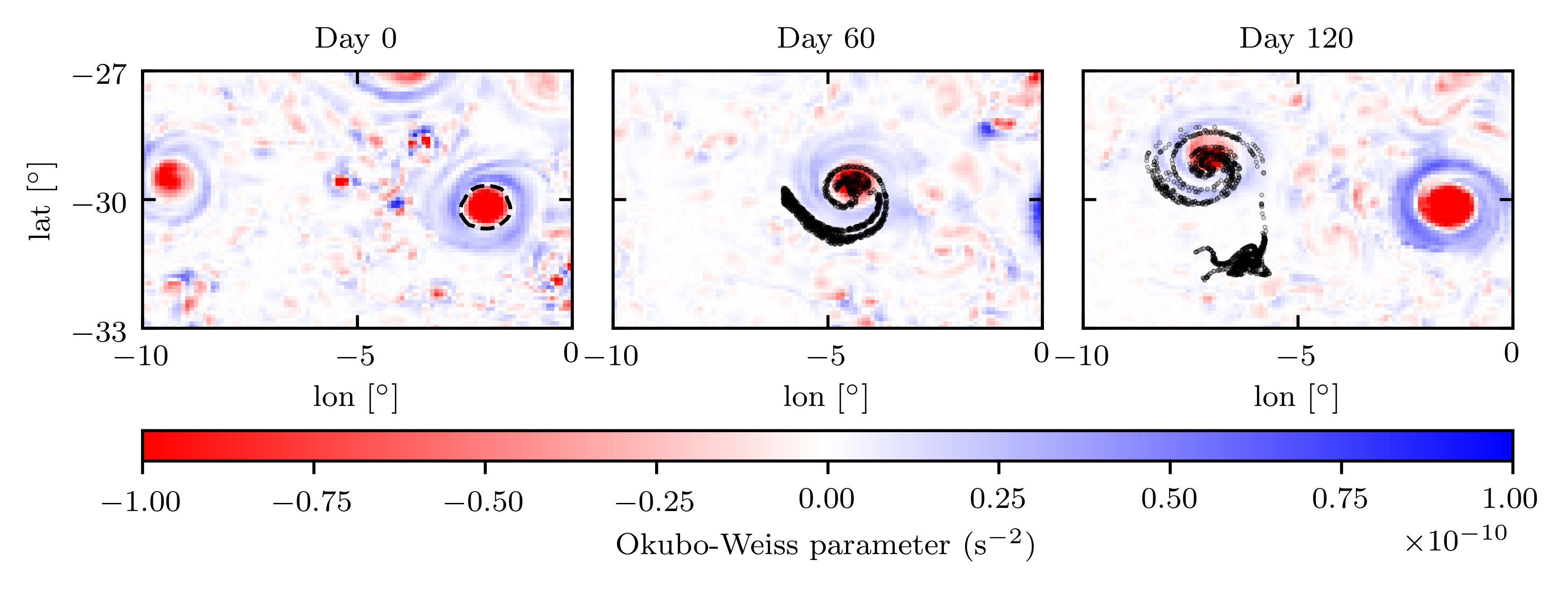}
\caption{Snapshots of the Okubo-Weiss parameter calculated using surface velocities from a global numerical ocean simulation. Red (blue) indicates regions where vorticity (strain) dominates. Virtual particles are initialized in a vorticity-dominated region in the first snapshot (indicated by the dashed black contour) and  advected using the surface velocity field. A substantial portion of particles (indicated by black dots in the second and third snapshots) are ejected from the eddy core while some remain entrained in the extended outer ring.}
\label{fig:intro_okubo_weiss}
\end{figure}

To illustrate the contrasting roles played by the eddy core and outer ring in transporting fluid, we show in Figure \ref{fig:intro_okubo_weiss} the evolution of an isolated mesoscale ocean eddy derived from daily surface velocities from a global numerical ocean simulation (model details are described in Section \ref{sec:data_numerics}). Three successive $60$-day snapshots of the Okubo-Weiss field in a region of the South Atlantic ocean clearly show the signature of vorticity-dominated regions in deep red. Virtual particles are initialized in the vorticity-dominated region in the first snapshot (indicated by the dashed black contour on the right side of the first snapshot) and advected horizontally using the 2D surface velocity field (the contribution of the vertical velocity is neglected).  While the Eulerian signature of the eddy is clearly visible as it propagates westward over the 120-day period, a substantial portion of the particles are ejected from the core in a long filament that detaches from the eddy to the south. The remaining particles are trapped within the core or become entrained in the outer ring (indicated by the shear-dominated blue region around the core) where they mix with material from the surrounding waters on shorter timescales.

This example and the earlier studies described above suggest that ocean eddies exhibit multiple timescales: a longer \emph{persistence time} or Eulerian lifetime, over which an eddy-like feature can be distinguished and tracked, and  multiple shorter \emph{material coherence timescales} or Lagrangian timescales that range from short (in the quasi-coherent outer ring) to potentially up to the Eulerian lifetime (for the long-lived coherent inner core).
Situated between these two extreme timescales is the median residence timescale, where half of the fluid has drained from the eddy feature. At this timescale, the eddy-like feature is still distinguishable, but has deviated from very strongly materially coherent motion.
In this paper, we examine material coherence in ocean eddies through a systematic study of the role played by the flow duration (or window length) over which fluid parcels are tracked. Flow duration is an important parameter for LCS studies: a long flow duration will help identify features that resist filamentation over long periods, and acts to filter out weakly coherent or leaky structures. 
For example, \cite{Froyland2014a} examined the effect of flow duration on finite-time coherent sets (as defined in \cite{Froyland2013}) and showed that with increasing flow duration, the boundary of the coherent set at the initial time will increasingly elongate, aligning with a stable manifold of the time-evolving flow.
When diffusion is added continuously in time, this alignment continues until it reaches an optimal balance between diffusion and advection.

There have been relatively few studies on the effect of flow duration to identifying ocean eddies. \cite{Wang2016} found that increasing flow duration reduced the ability of coherent material loops to resist stretching, thereby shrinking the area enclosed by the loop to a `hard core' that remains stable over the Eulerian lifetime of the eddy. 
\cite{Abernathey2018}, using the LAVD method \cite{Haller2016}, found that there were far fewer identified coherent inner cores for long flow times ($90$ days) than shorter flow times ($30$ days), and effectively no coherent inner cores for flow times of $270$ days or greater in their study region.
\cite{AndradeCanto2020} develop a life expectancy by searching for the largest flow time that a rotationally-coherent loop \cite{Haller2015} is still successfully identified.
However, the lifetimes identified in each of these studies are those of a materially coherent inner core, and to date there has been no study of the material coherence timescales associated with an eddy's quasi-coherent outer ring.

To explore these issues, we will use the theory of finite-time coherent sets, originally developed in \cite{Froyland2010,Froyland2013}. 
The geometric theory of finite-time coherent sets \cite{Froyland2015a} seeks regions of phase space whose evolving boundary length to enclosed volume remains small over a specified time window.
The small evolving boundary length is directly related to reducing small-scale mixing in the presence of low-level diffusion.
These regions can be found by considering level sets of the dominant eigenfunction of a dynamic Laplace operator \cite{Froyland2015a} defined on this time window.
We study the effect of both window length (flow duration), and window center time (flow initialization time), on the computed eigenfunctions and selected level sets.
These effects naturally allow for the quantification of transport by both an eddy's coherent inner core as well its quasi-coherent outer ring. 

The dominant eigenvalue of the dynamic Laplace operator with Dirichlet boundary conditions describes the amount of mass loss (or, with Neumann boundary conditions, the amount of mixing), and so we introduce a new heuristic quantity, the \emph{maximal coherence timescale}, which minimizes the rate of mass loss per unit flow time over a domain. This can be used to determine a typical Lagrangian timescale of an ocean eddy. 
We use this maximal coherence timescale to study the quasi-coherent outer ring of an ocean eddy. More precisely, we examine the persistence and material coherence of this quasi-coherent outer ring by considering a sequence of finite-time coherent sets, parameterized by the window center time. 
We find that the eddy in this study exhibits no long-lived inner core, and that the transport statistics (described by residence time statistics) of the quasi-coherent outer ring indicate a significant ability to transport surface water. Additionally, these transport statistics are robust to the choice of window length.

The paper is organized as follows. Section \ref{sec:theorymethod} begins by providing a brief overview of the dynamic isoperimetric problem and its approximate solution derived from the associated dynamic Laplace operator. 
In section \ref{sec:windowsequence} we define two sequences of coherent sets.
The first of these sequences is a  \emph{telescoping sequence} constructed by fixing the window center time and expanding the window length, and the second is a \emph{sliding sequence} constructed by fixing the window length and translating the window center time. 
Using the telescoping sequence of coherent sets, we introduce a maximal coherence timescale, which provides a timescale over which a finite-time coherent set experiences a minimal mass loss per unit flow time. Section \ref{sec:data_numerics} introduces the dataset and numerical methods we will use to study the dominant eigenvalues and eigenfunctions of the two sequences of finite-time coherent sets.
Section \ref{sec:application} contains a case study of finding a typical maximal coherence timescale of an Agulhas ring, and a case study of understanding the material coherence of a decaying Agulhas ring. By considering the residence times of particles contained in a particular sequence of finite-time coherent sets, we show our method can identify an eddy even if no long-lived coherent inner core exists. We show that the residence time statistics are robust to the choice of window length, and that the quasi-coherent outer ring of the eddy significantly contributes to the transport of surface water by the eddy. We also show that the maximal coherence timescale depends on the window center time, as an eddy strengthens or weakens throughout its life. A discussion and conclusion is found in Section \ref{sec:discussion}.

\section{Theory and Method}\label{sec:theorymethod}
\subsection{Finite-time coherent sets and the dynamic Laplace operator}
\label{sec:DL}

Finite-time coherent sets are regions of phase space that minimally mix with the remainder of the flow domain over a finite interval of time.
They are regions whose boundaries best resist distortion and filamentation under finite-time evolution;  there is therefore minimal mixing due to small-scale diffusion at the interface of the boundary. 
The process of identifying finite-time coherent sets can be recast as a dynamic isoperimetric problem \cite{Froyland2015a,FroylandKwok}, which seeks sets with minimal evolving boundary size relative to enclosed volume.
In this section we briefly review the dynamic isoperimetric problem and its approximate solution using superlevel sets of the dynamic Laplace operator.
We are interested in features that do not intersect the boundary of the domain (e.g. an eddy in the open ocean), so we focus here on the Dirichlet boundary condition case \cite{Froyland2018}. 
The Neumann boundary condition case, which applies to features that may intersect the domain boundary is developed in \cite{Froyland2015a}.

\begin{figure}[h]
\centering
\includegraphics[width=0.6\linewidth]{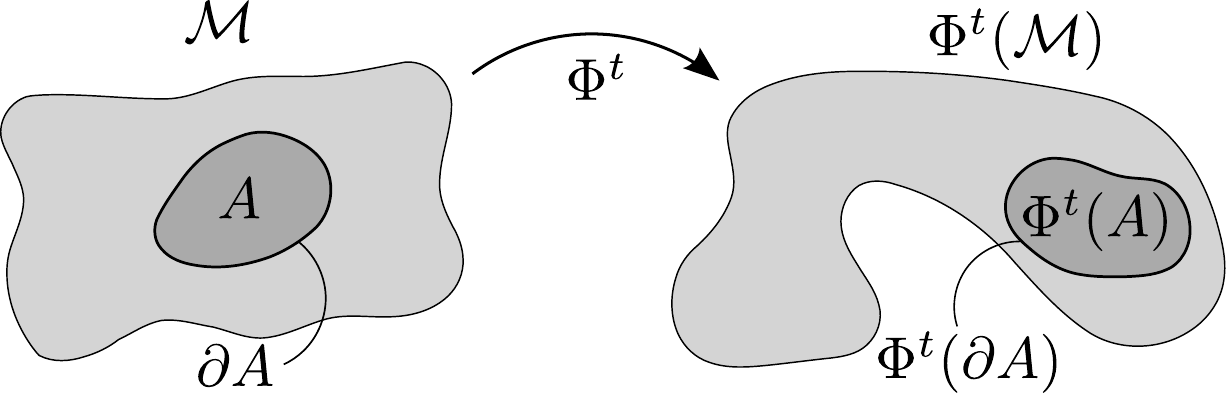}
\caption{Evolving submanifold $A$.  The boundary $\partial A$ of $A$ disconnects $A$ from the rest of $\mathcal{M}$ without intersecting the boundary $\partial \mathcal{M}$ of $\mathcal{M}$.}
\label{fig:evolving_manifold}
\end{figure}

Given a volume-preserving flow and a compact smooth Riemannian manifold $\mathcal{M}\subset \mathbb{R}^d$, we seek a submanifold $A \subset \text{int}(\mathcal{M})$ whose average evolving boundary is minimal relative to its volume.
Such a set $A$ is a natural candidate for a finite-time coherent set since it minimizes the length of the interface over which small-scale diffusive mixing can occur (Figure \ref{fig:evolving_manifold}).
For $t\in \mathbb{R}$ we consider smooth flow maps $\Phi^{t} : \mathcal{M} \rightarrow \Phi^{t}(\mathcal{M})$ describing evolution in phase space from time $0$ to time $t$.
We also define corresponding evolutions on functions:  let the push-forward $\Phi^t_*:C^{\infty}(\mathcal{M}) \rightarrow C^{\infty}(\Phi^{t}(\mathcal{M}))$ be defined by $\Phi^t_*f=f\circ\Phi^{-t}$, and the pull-back $(\Phi^{t})^{*} : C^{\infty}(\Phi^{t}(\mathcal{M})) \rightarrow C^{\infty}(\mathcal{M})$ be
defined by $(\Phi^t)^*f=f\circ\Phi^{t}$.

We seek the evolving submanifold $\Phi^t (A)$ that is minimally distorted over the time interval or \emph{window} $W(0,T) = [-T/2,T/2]$ of length $T>0$ centered at time $t = 0$.
We define a version of the dynamic Cheeger constant \cite{Froyland2015a,Froyland2018} over $W(0,T)$ 
as
\begin{equation} \label{dynamic_cheeger}
\mathbf{h}^D(A) := \frac{\frac{1}{T}\int_{-T/2}^{T/2} \ell_{d-1}(\Phi^{t}(\partial A)) \text{ d}t} {\ell(A)},
\end{equation}
where $\ell$ is $d$-dimensional volume and $\ell_{d-1}$ is codimension 1 volume. 
This is the time-averaged surface measure of the evolved hypersurfaces $\Phi^{t}(\partial A)$ over $t \in W(0,T)$, normalized by the volume of $A$. 
The submanifold $A$ that minimizes the ratio (\ref{dynamic_cheeger}) is the solution to our dynamic isoperimetric problem, and we denote this minimal ratio by $\mathbf{h}^D$:
\begin{equation} \label{min_dynamic_cheeger}
\begin{split}
\mathbf{h}^D := \text{min}\{\mathbf{h}^D(A) : A &\text{ is an open submanifold of int}(\mathcal{M}),\\
&\text{with compact closure and } C^{\infty}\text{ boundary}\}.
\end{split}
\end{equation}
From the dynamic Federer-Fleming Theorem \cite{Froyland2015a}, we can express $\mathbf{h}^D$ as a functional minimization problem
\begin{equation} \label{dynamic_sobolev}
\mathbf{h}^D = \mathbf{s}^D := \underset{f \in C^{\infty}_{c}(\mathcal{M},\mathbb{R})}{\text{inf}} \frac{\frac{1}{T}\int_{-T/2}^{T/2}||\nabla(\Phi^{t}_{*}f)||_{1} \text{ d}t}{ ||f||_{1}},
\end{equation}
where $\mathbf{s}^D$ is the dynamic Sobolev constant from \cite{Froyland2015a}, and $C^{\infty}_{c}(\mathcal{M},\mathbb{R})$ denotes the set of non-identically vanishing, compactly supported, $C^{\infty}$ real-valued functions on int$(\mathcal{M})$.

The optimal value $\mathbf{s}^D$ of the $L^1$-minimization problem (\ref{dynamic_sobolev}) is bounded by the optimal value of a simpler, analogous $L^2$-minimization problem. Specifically, we define the \emph{dynamic Laplace operator} \cite{Froyland2015a}
\begin{equation}\label{dynamic_laplace}
\Delta^{D}_{W(0,T)} := \frac{1}{T}\int_{-T/2}^{T/2} (\Phi^{t})^{*} \Delta \Phi^{t}_{*}\ dt,
\end{equation}
with associated Dirichlet eigenproblem
\begin{equation}\label{dirichlet_eigenproblem}
\Delta^D_{W(0,T)} v = \lambda v  \quad \text{on int}(\mathcal{M}) \qquad \text{with }  
v = 0 \quad \text{on } \partial{\mathcal{M}}.
\end{equation}

For later use, we state a weak form of (\ref{dirichlet_eigenproblem}).
Multiplying (\ref{dirichlet_eigenproblem}) by a test function $\psi:\mathcal{M}\to\mathbb{R}$, integrating over $\mathcal{M}$, expanding $\Delta^{D}_{W(0,T)}$ as in (\ref{dynamic_laplace}), and using Green's formula, one arrives at a weak form of the eigenproblem (derived in \cite{Froyland2018}):  find eigenpairs $(\lambda, v)$, $v\in H_0^1(\mathcal{M})$ such that
\begin{equation} \label{weak_eigenproblem}
-\frac{1}{T} \int_{-T/2}^{T/2} \int_{\Phi^{t}(\mathcal{M})} \nabla (\Phi^{t}_{*} v) \bullet \nabla (\Phi^{t}_{*} \psi) \text{ d}l\text{d}t = \lambda \int_{\mathcal{M}} v\psi \text{ d}l \text{ for all }\psi \in H^{1}_{0}(\mathcal{M}), 
\end{equation}
where $H_0^1(\mathcal{M})$ is the space of compactly supported complex-valued functions on $\mathcal{M}$ with square-integrable derivative.

Returning to the eigenproblem (\ref{dirichlet_eigenproblem}), the leading eigenvalue of $\Delta^D_{W(0,T)}$ is
	\begin{equation} \label{variatonal_characterisation}
	\lambda = -\underset{v \in C^{\infty}_{c}(\mathcal{M},\mathbb{R})}{\text{inf}} \frac{\frac{1}{T}\int_{-T/2}^{T/2}||\nabla(\Phi^{t}_{*}v)||_{2}^2 \text{ d}t}{ ||v||_{2}^2},
	\end{equation}
with the infimum achieved only when $v$ is the eigenfunction corresponding to $\lambda$.
The optimal value $\mathbf{h}^D$ of the $L^{1}$ minimization is bounded above using  the dynamic Cheeger inequality \cite{Froyland2015a}:
$\mathbf{h}^D \le 2\sqrt{-\lambda}$. Similarly, the dynamic Buser inequality \cite{Rock2021} provides a lower bound: $\mathbf{h}^D \ge b \sqrt{- \lambda}$ where $0 < b < 2$ is a constant determined by the dimension and Ricci curvature of the manifold, and the maximum amount of stretching under the dynamics. 
In summary,
\begin{equation} \label{h_bounds}
b \sqrt{-\lambda} \le \mathbf{h}^D \le 2\sqrt{-\lambda}.
\end{equation}
The eigenfunction $v$ is strictly positive in the interior of $\mathcal{M}$ by the maximum principle \cite{Olver2014}, and is a spatially smoothed version of the minimizing $f$ in equation (\ref{dynamic_sobolev}).

The boundary of the corresponding finite-time coherent set $A$ can be approximated by a level set of $v$ \cite{Froyland2015a,Froyland2018}. 
Let $A_{a} = \{x \in \mathcal{M} : v(x) \ge a\}$ denote a superlevel set of $v$ at value $a$.
Note that the eigenvector $v$ has been computed over the window $W(0,T)$ and is anchored at time 0.
 
\subsection{Sequences of coherent sets} \label{sec:windowsequence}
\begin{figure}[h]
\centering
\begin{subfigure}{0.5\linewidth}
	\centering
	\includegraphics[width=\linewidth]{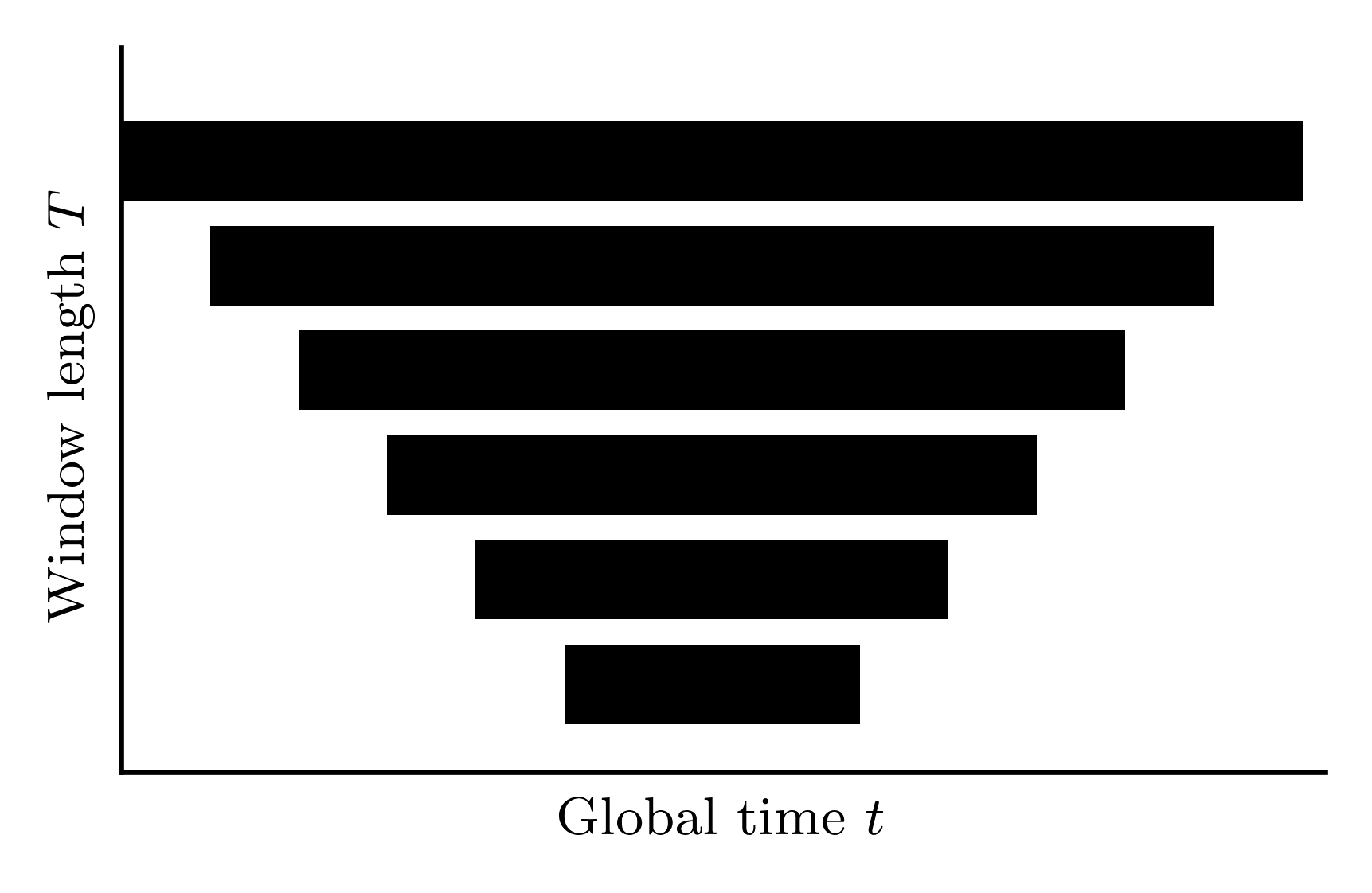}
	\caption{telescoping windows}
	\label{fig:telescoping_window}
	
\end{subfigure}%
\begin{subfigure}{0.5\linewidth}
	\centering
	\includegraphics[width=\linewidth]{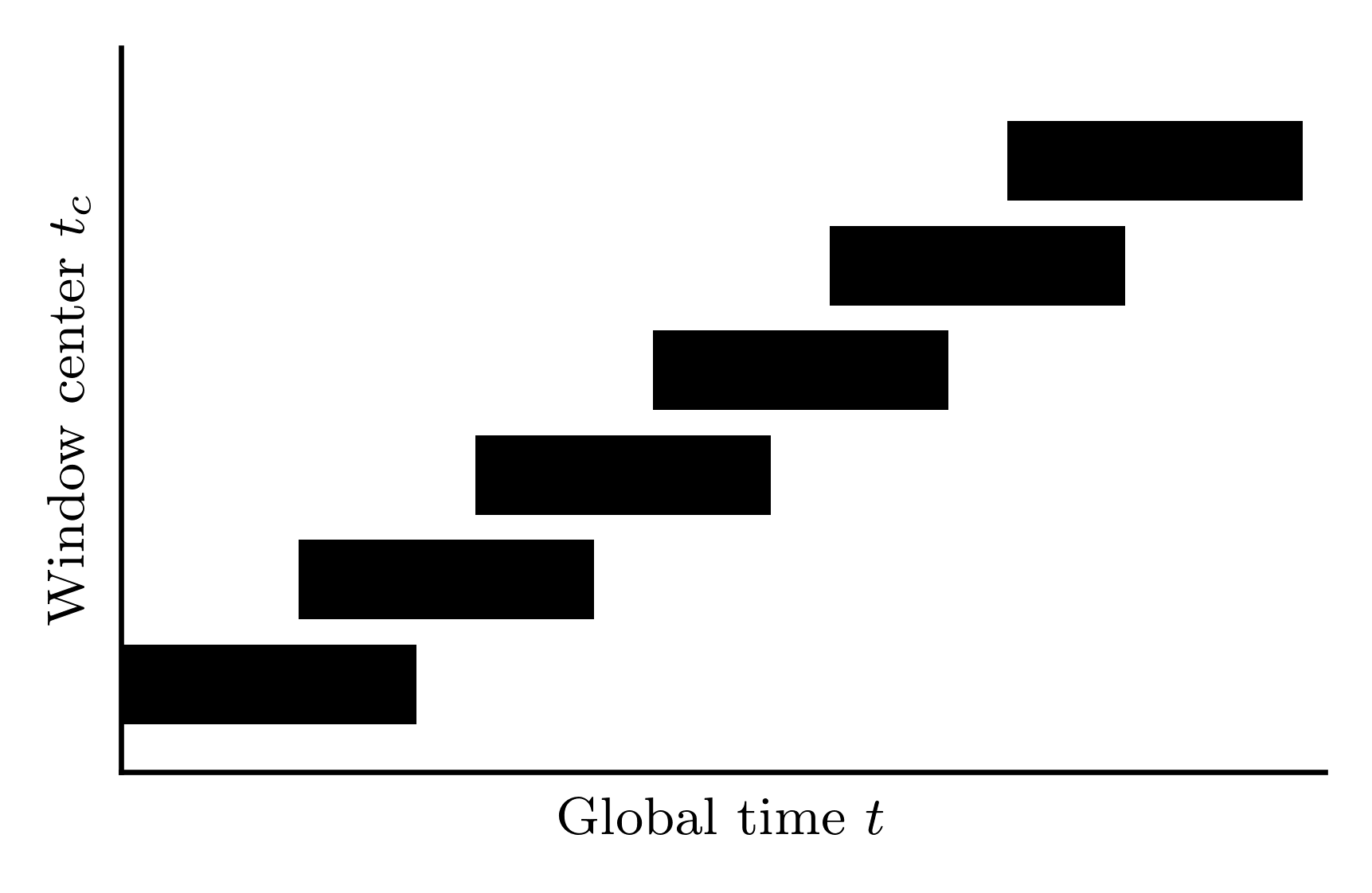}
	\caption{sliding windows}
	\label{fig:sliding_window}
\end{subfigure}
\caption{Visual representations of two approaches to creating sequences of finite-time coherent sets.} 
\label{fig:windowing_approach}
\end{figure}

The coherent sets identified in the previous subsection depend upon the choice of time window.
We can vary the window width $T$ and center time $t_c$, leading to a two-parameter family of windows $W(t_c,T):=[t_c-T/2,t_c+T/2]$.
We construct corresponding \emph{sequences of coherent sets} by systematically varying one or both of these parameters. 
Although the resulting parameterized family of sets is referred to here as a sequence, the variation of $T$ and $t_c$ is, in principle, continuous. Differentiable changes to the window are expected to lead to differentiable changes in the resulting coherent set \cite{AFJ21}.
In practice, we numerically construct the sequence using a discrete set of parameter values for $T$ and $t_c$.
We consider two sequences of coherent sets (see also Figure \ref{fig:windowing_approach}):

\textbf{Telescoping sequence:} 
by fixing $t_c$ and varying $T$, we may construct a \emph{telescoping sequence} of coherent sets based on the windows $W(t_c,T)$.
For chosen $T_{\min}$, and $T_{\max}$, such that $T_{\min} < T_{\max}$, we have a sequence of windows $\{W(t_c,T)\}_{T\in[T_{\min},T_{\max}]}$ from which we can construct a telescoping sequence of coherent sets.

\textbf{Sliding sequence:}
by fixing $T$ and varying $t_c$, we may construct a \emph{sliding sequence} of coherent sets based on the windows $W(t_c,T)$.
For chosen $t_{\min}$, and $t_{\max}$, such that $t_{\min} < t_{\max}$, we have a sequence of windows $\{W(t_c,T)\}_{t_c\in[t_{\min},t_{\max}]}$ from which we can construct a sliding sequence of coherent sets.

Telescoping and sliding sequences can be used to reveal important characteristics of the coherent object being studied. For example, \cite{Froyland2015c} used a sliding sequence of finite-time coherent sets to study the decay of an ocean eddy. More recently, \cite{schneide2021} used a sliding sequence of windows to study the Lagrangian pathways of convective heat transfer in turbulent Rayleigh-B\'enard convection flows. 
Sliding sequences of windows have also been used to find structural changes in coherent features \cite{blachut2020,blachut2021,ndour2021}, such as merging or splitting. The effect of telescoping windows on the geometry of finite-time coherent sets was investigated in \cite{Froyland2014a} and  \cite{AndradeCanto2020} have recently used a telescoping sequence of windows to determine the life expectancy of an ocean eddy by finding the largest window length that successfully identified a coherent vortex.  
An alternative approach to detecting the birth and death of finite-time coherent sets was developed in \cite{Froyland2021}, where an inflated dynamic Laplace operator on a time-expanded domain removes the need for multiple telescoping and sliding computations.

Longer windows will lead to sliding sequences of coherent sets that are more material and therefore better transporters of fluid.  
On the other hand, longer windows will reduce the total sequence length over which a coherent feature can be identified. 
This is consistent with the properties of Eulerian estimators like the Okubo-Weiss parameter, which can provide very long sequences of eddy-like features that are not particularly good material transporters of fluid.
By varying the window length, one can identify a \emph{maximal coherence timescale} over which the mass loss (or mixing, with Neumann boundary conditions) per unit time by an eddy is minimized. Likewise, by varying the center time and examining the change in particles within overlapping coherent sets, one can quantify the rate of leakage or entrainment of material by an eddy, and capture the growth, decay, and \emph{quasi-coherence} of an eddy. These topics are taken up in sections \ref{sec:telescoping}, and \ref{sec:tracking}.

\subsection{A maximal coherence timescale} \label{sec:2.3}
For a fixed $t_c$, systematic variation of the window length $T$---that is, a telescoping sequence of windows---can be used to identify a window length $\hat{T}$, which we call the \emph{maximal coherence timescale}.
Suppressing the $t_c$ dependence, let $\lambda(T)$ denote the leading eigenvalue of $\Delta^D_{W(t_c,T)}$.
We define this maximal coherence timescale $\hat{T}$ as 
\begin{equation} 
\label{optimal_t}
\hat{T} := \underset{T>0}{\mbox{arg max }} \frac{\lambda(T)}{T},
\end{equation}
under Dirichlet boundary conditions\footnote{If one has Neumann boundary conditions, $\lambda$ is interpreted as the leading nontrivial (nonzero) eigenvalue.}.
The quantity $\lambda(T)/T$ can be interpreted as the rate of mass loss\footnote{In the Neumann boundary condition case we would interpret $\lambda(T)/T$ as the rate of mixing of the eigenfunction $v$ per unit window length $T$.} per unit window length $T$.
In Appendix \ref{sec:timescale} we elaborate on this interpretation and connect it with an equivalent maximal coherence timescale computed instead with  singular values of transfer operators, following \cite{Froyland2013}.
For the moment, let us consider some useful simple cases.

\textbf{Chaotic flow:} If the dynamics completely lacks coherence, we expect $\lambda(T)/T$ to rapidly become more negative as $T$ increases.  
This is because it is impossible to ``quarantine'' mass away from the boundary where it will leak from the domain\footnote{In the Neumann case, there is no mass loss, but it is not possible to isolate positive mass from negative mass in the eigenfunction $v$ and they are rapidly mixed as $T$ increases.} due to the Dirichlet boundary conditions, and thus $\lambda(T)$ rapidly decreases. 
Because of finite data resolution, for large enough $T$ the rate of decrease of $\lambda(T)$ will slow, and at some finite $T>0$ the quotient $\lambda(T)/T$ will achieve its minimum (the least coherent timescale relative to the data resolution).
The quotient $\lambda(T)/T$ will then increase slowly as $T$ further increases, but never exceeding its initial value.
In summary, the maximal coherence timescale is very short, consistent with chaotic flow.

\textbf{Coherent flow:} If the domain contains a single strongly coherent set we may increase $T$ without decreasing $\lambda(T)$ appreciably.
This is because we may isolate most of the mass inside the coherent set, preventing it from moving to the boundary and leaking out\footnote{In the Neumann case, positive mass may be isolated in the coherent set and prevented from mixing with negative mass outside the coherent set.}.
With $\lambda(T)$ relatively stable and $T$ increasing, we expect a monotonic increase in the quotient $\lambda(T)/T$, achieving its maximum at arbitrarily large $T$.
In other words the maximal coherence timescale is very large, consistent with coherent flow.

\textbf{Partially coherent flow:}
In this setting, which models our eddy dynamics, the behavior of $\lambda(T)/T$ can be more complex than the two above simpler cases.
Because on short timescales our flow contains coherent dynamics (i.e.\ our eddy remains coherent), we can expect an initial increase in the quotient $\lambda(T)/T$.
As we increase $T$ further, the partially coherent object (e.g.\ our eddy) begins to break up and fails to be completely coherent.
At some optimal window length $\hat T$ the benefit we initially obtained from short-term coherence is outweighed by break-up induced by longer windows, and the quotient $\lambda(T)/T$ will stop increasing, turn over, and begin to decrease.
At this turning point $\hat{T}$, we reach the maximum of $\lambda(T)/T$ where our rate of mass loss\footnote{In the Neumann boundary condition case our rate of mixing is least per unit window length.} is least per unit window length.
In summary, the maximal coherence timescale is finite.

\begin{figure}[h]
\centering
\includegraphics[width=1\linewidth]{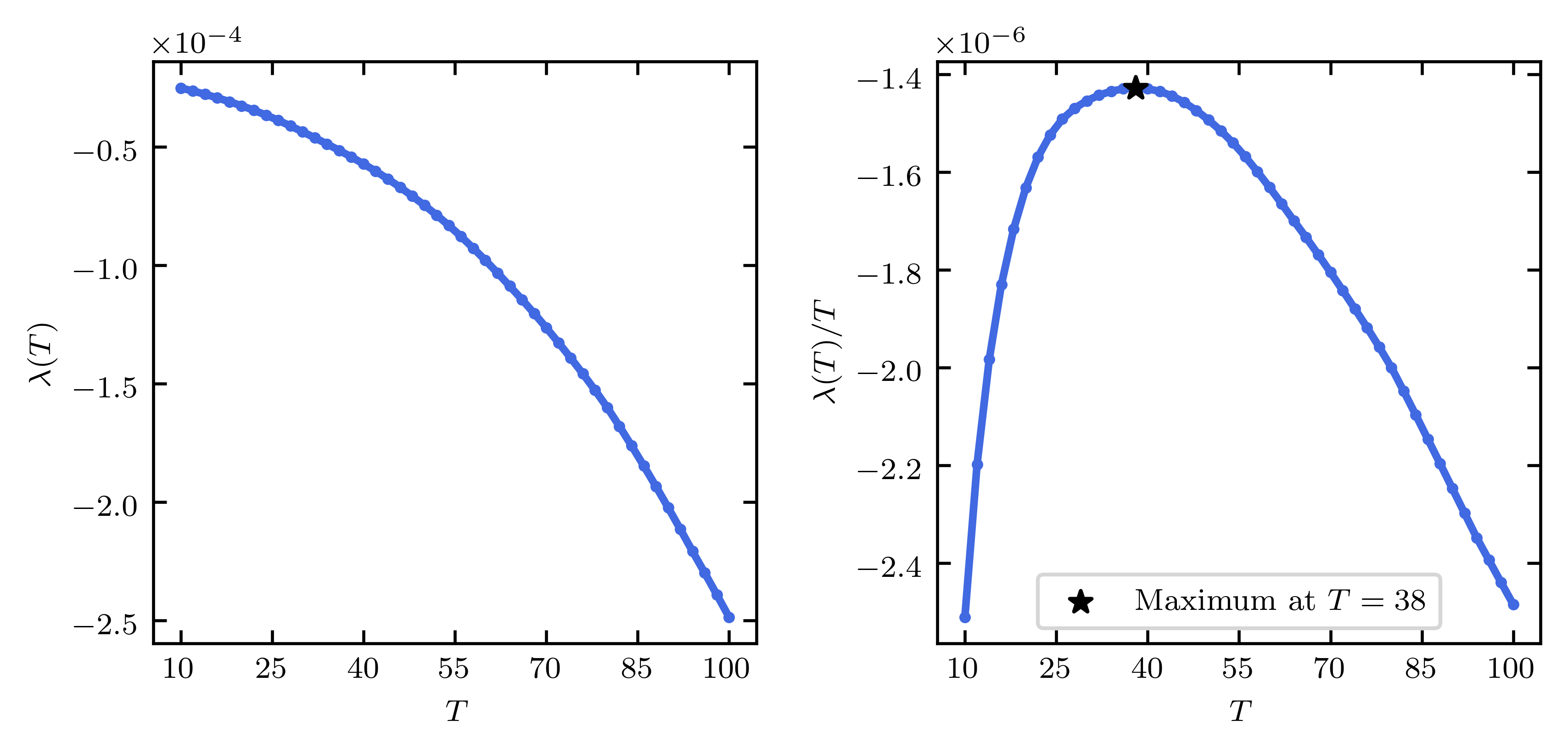}
\caption{(Left) Dominant eigenvalue $\lambda(T)$ of $\Delta^{D}_{W(50,T)}$
for a range of window lengths $T$. As $T$ increases, $\lambda(T)$ monotonically decreases indicating increasing \emph{total} mass loss across the window of duration $T$. (Right) Ratio $\lambda(T)/T$ quantifying the mass loss per-unit-window-duration. As described in Section \ref{sec:telescoping}, we see a maximum of this ratio at $\hat T=38$, which we call the maximal coherence timescale for the eddy at $t_c=50$.}
\label{fig:window_50_eigenvalues}
\end{figure}

In practice, the maximization in (\ref{optimal_t}) is performed over a sequence of windows $\left\{ W(t_c,T)\right\}_{T \in \left[ T_{\mathrm{min}}, T_{\mathrm{max}} \right]}$.
To illustrate the expected behavior for partially coherent flow, Figure \ref{fig:window_50_eigenvalues} (Left) plots $\lambda(T)$ vs. $T$ and
Figure \ref{fig:window_50_eigenvalues} (Right) shows $\lambda(T)/T$ vs. $T$, for the eddy we analyze in Section \ref{sec:application}.
In section \ref{sec:decay} we show that as $t_c$ varies, $\hat{T}$ may also change, and so $\hat T$ is a function of $t_c$.

\section{Data and Numerics}\label{sec:data_numerics}

\subsection{Ocean model and trajectory integration} \label{sec:data_domain_traj}
We use surface velocity data
derived from the  Australian Community Climate and Earth System Simulator global ocean-sea ice model (ACCESS-OM2) \cite{Kiss2020}. 
The model couples the MOM5.1 ocean model to the CICE5.1.2 sea ice model via OASIS3-MCT and is forced using prescribed atmospheric conditions from the JRA55-do reanalysis. We use daily simulation output from the ACCESS-OM2-01 ``eddy-rich'' configuration of the model, which has a horizontal resolution of $0.1^\circ$ between $65^\circ$S and $65^\circ$N and 75 vertical levels and fully resolves the active mesoscale eddy field in most of the ocean \cite{Kiss2020}.

Our focus in this study is a region in the South Atlantic where Agulhas rings regularly form, evolve, and decay, denoted $\mathcal{M}$, which may be represented in polar coordinates as $[25^\circ$W$, 20^\circ$E$]\times[40^\circ$S, $20^\circ$S$]$. 
In this domain, particles are initialized at the ocean surface with a uniform spacing of $0.025^\circ$ in  longitude and latitude coordinates, resulting in $1,440,000$ particles.
Particles are advected forward in time using daily ocean surface velocity fields and the Lagrangian ocean analysis tool Parcels \cite{Lange2017,Delandmeter2019} from the model date 30/12/2009 for $730$ days with a $5$-minute timestep using the Runge-Kutta advection scheme.
As the domain partially covers land, we remove approximately 114,000 particles that either begin on or are advected onto land during the study period, resulting in a total of $n = 1,326,117$ valid particle trajectories.

The particle positions $x_{i}^{t} \in \mathcal{M}$ are stored as longitude-latitude coordinates $(\alpha_i^{t},\beta_i^{t})$ where $i \in \{1,\dots,n\}$ is the particle index and $t \in \tau = \{0,1,\dots,730\}$ are observation times measured in days from release. The trajectories are used to approximate flow maps in this region over the window $W(0,730)$. We denote by $\Phi_0^t$ the flow map generated by particles released on day 0 and advected for $t$ days, i.e., $x_i^t = \Phi_0^t x_i^0$ where $x_i^0$ is the initial position of the $i$th particle. The initial time of the flow map will be an important parameter in our study and so we introduce a modified notation. Specifically, the flow map from time $t_c$ to time $t_c + t$ will be denoted as
\begin{equation}\label{Phi_tc}
\Phi_{t_c}^t=\Phi^{t_c+t}_0\circ (\Phi^{t_c}_0)^{-1}, 
\end{equation}
that is, particles are evolved backwards in time from $t_c \ge 0$ to day 0 (when they are on a uniform grid), then forwards in time to $0 \le t_c + t \le 730$, where $t$ can take positive or negative values. In practice we simply evolve trajectories forwards or backwards from time $t_c$ to $t_c + t$.

\subsection{Domain selection}\label{sec:domain}

We restrict the physical domain to study the evolution of a single Agulhas eddy. The domain is selected by first identifying a target eddy from the surface velocity field on day 0, and second defining a bounding longitude-latitude disc that completely encloses the eddy as it evolves in time.

An initial survey of the surface velocity field was performed by calculating the Okubo-Weiss field of the flow \cite{Okubo1970,Weiss1991}, 
\begin{equation}\label{OW-param}
    OW = \left( \frac{\partial u}{\partial x} - \frac{\partial v}{\partial y} \right)^2 + \left( \frac{\partial v}{\partial x} + \frac{\partial u}{\partial y} \right)^2  - \left( \frac{\partial v}{\partial x} - \frac{\partial u}{\partial y} \right)^2, 
\end{equation}
where $x, y$ are the eastward and northward coordinates (measured in meters) and $u, v$ are eastward and northward velocities (in meters per second). $OW$ is the sum of the squares of the normal and shear strain minus the vorticity squared and is a measure of the relative importance of deformation and rotation at each location. Negative values of $OW$ indicate regions where rotation dominates strain.

$OW$ was calculated from gridded surface velocity fields for model date 30/12/2009 (day 0) with a finite difference scheme using the model grid spacing, thereby accounting for the curvature of Earth's surface.
By visual inspection, we selected one closed vorticity-dominated region with $OW \le -2\times10^{-12} \text{s}^{-2}$, a standard threshold for identifying ocean eddies (e.g. \cite{Chelton2007}). This is an initial approximation of the eddy location on this date, which we denote as $Z(0)$. For subsequent times $t_i \in \tau$ we compute $OW$ using the surface velocity fields for day $t_i$. The eddy is tracked by computing the centroid of each vorticity-dominated region, then defining $Z(t_i)$ as the vorticity-dominated region with centroid closest to the centroid of $Z(t_{i-1})$. This produces a sequence $\{Z(t)\}_{t\in\tau}$ of vorticity-dominated closed regions that approximate the location of the eddy at each timestep.

The neighborhood of the eddy is approximated by a fixed radius disc that completely encloses the eddy during a given time window. For each window $W(t_c,T)$ we have a discrete set of (daily) times $\overline{W}(t_c,T) := W(t_c,T)\cap\tau$.
Define $\mathbf{Z}_{t_c,T}$ to be the region containing the push-forwards and pull-backs of the approximate eddy locations over this period to the center time $t_c$,
\begin{equation*}
\mathbf{Z}_{t_c,T} = \bigcup_{t \in \overline{W}(t_c,T)} \Phi^{t_c-t}_{t} Z(t),
\end{equation*}
and define by $c_{t_c,T}$ the centroid of $\mathbf{Z}_{t_c,T}$.
We define a subdomain $\mathcal{M}_{t_c,T} \subset \mathcal{M}$ as a bounding disc defined by $\{x^{t_c} \in \Phi^{t_c}_{0}(\mathcal{M}) : d(x^{t_c},c_{t_c,T}) \le r \}$, where $d(x,y)$ is the great-circle distance between points $x$ and $y$, and $r$ is a radius selected large enough to completely contain $\mathbf{Z}_{t_c,T}$.

To numerically approximate $\mathcal{M}_{t_c,T}$, we first approximate the the push-forwards and pull-backs of $Z(t)$ by finding the indices of particles contained inside $Z(t)$, that is $\{i: x_{i}^{t} \in Z(t)\}$. By taking the union of these sets over the discrete window $\overline{W}(t_c,T)$, we define the indices $S_{t_c,T}$ of particles in $\mathbf{Z}_{t_c,T}$ as
\begin{equation}\label{vorticity_domain_indices}
S_{t_c,T} := \bigcup_{t\in \overline{W}(t_c,T)} \{i: x_{i}^{t} \in Z(t)\}.
\end{equation}
Because $S_{t_c,T_1}\subset S_{t_c,T_2}$ for $T_1<T_2$, in our experiments in Section \ref{sec:telescoping} we will set $\mathcal{M}_{t_c,T} = \mathcal{M}_{t_c,T_{\max}}$ to make consistent comparisons for differing $T$.

We approximate $\mathbf{Z}_{t_c,T}$ by $\{x^{t_c}_{i}\}_{i\in S_{t_c,T}}$, and compute $c_{t_c,T}$ by taking an average of the longitudes and latitudes of particles contained in $\mathbf{Z}_{t_c,T}$.
We define  $I_{t_c,T} := \{i: d(x_{i}^{t_c},c_{t_c,T}) \le r \}$ as the indices of particles contained in $\mathcal{M}_{t_c,T}$, and estimate $\mathcal{M}_{t_c,T}$ as $\bigcup_{i\in I_{t_c,T}} x_{i}^{t_c}$. 

\begin{figure}
\centering
\includegraphics[width=1\linewidth]{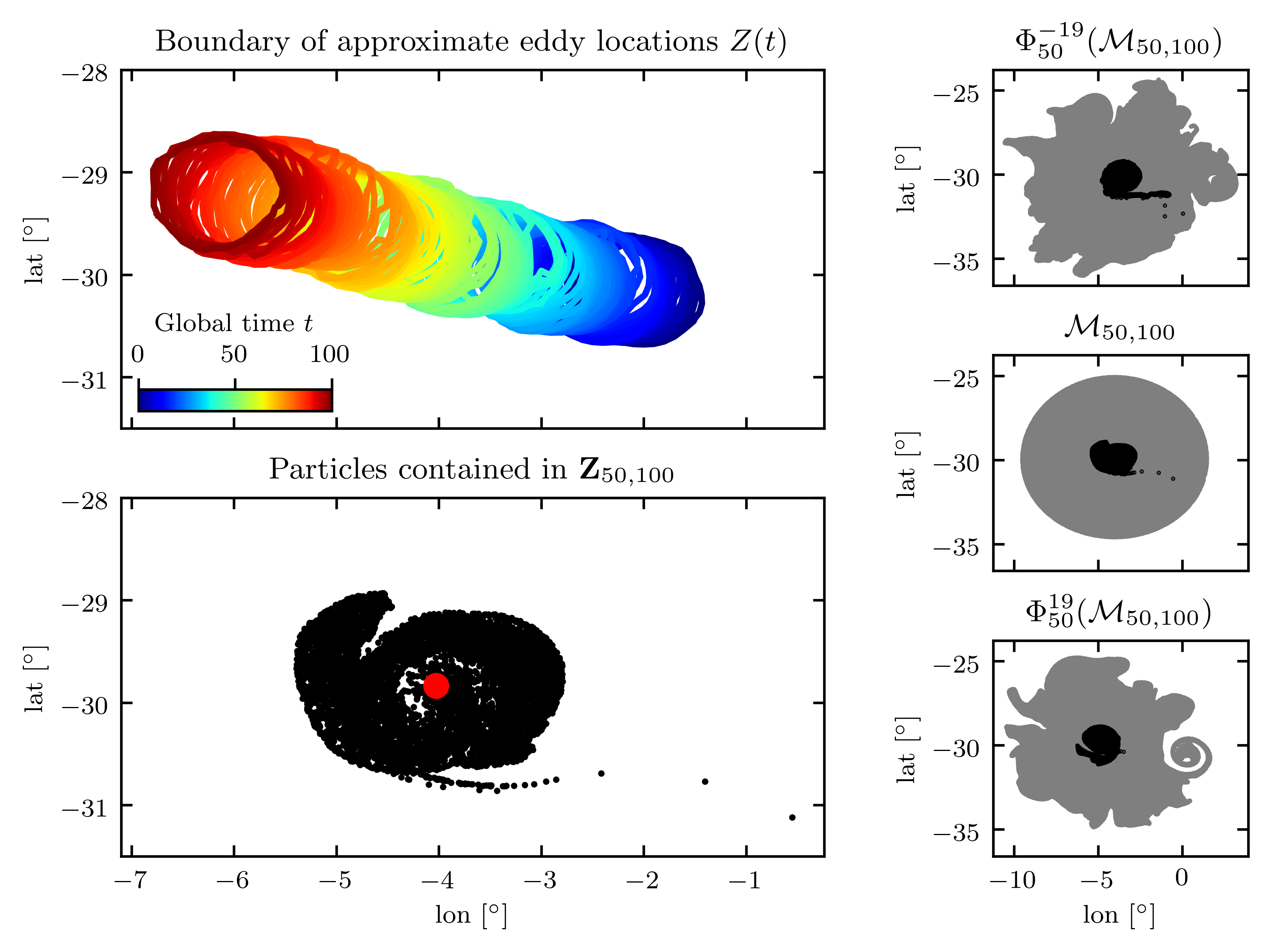}
\caption{(Left Top) Plots of boundaries of $Z(t)$ for $t \in \{0,1,\ldots,100\}$, colored according to global time $t$. (Left Bottom) Particles contained in $\mathbf{Z}_{50,100}$ and the centroid plotted in red. (Right Middle) Particles contained in $\mathcal{M}_{50,100}$ (black and gray), and $\mathbf{Z}_{50,100}$ (black), along with their pull-backs by 19 days (Right Top), and push-forwards by 19 days (Right Bottom).}
\label{fig:example_domain}
\end{figure}

The process for selecting the subdomain $\mathcal{M}_{50,100}$ is illustrated in Figure \ref{fig:example_domain}.
Firstly, the target eddy is identified on day 0 and tracked forward in time to define the sequence $\{Z(t)\}_{t\in\tau}$. The boundaries of $Z(t)$ are shown for all $t \in \overline{W}(50,100)$, and the vorticity dominated region $\mathbf{Z}_{50,100}$ is shown at time $t=50$. We also plot the subdomain $\mathcal{M}_{50,100}$, along with its pull-back and push-forward by $19$ days, which shows highly nonlinear $\Phi^{t}$. 

\subsection{Mesh construction}\label{sec:triangulation}
We use a finite-element approach \cite{Froyland2018} to numerically approximate the dynamic Laplacian $\Delta^D_{W(t_c,T)}$.
This approach requires a triangular mesh of our domain $\mathcal{M}_{t_c,T}$ 
as well as meshes of the domains $\Phi^{t - t_c}_{t_c}(\mathcal{M}_{t_c,T})$ for $t \in \overline{W}(t_c,T)$.
Delaunay triangulation is an obvious choice of triangulation, however $\Phi^{t-t_c}_{t_c}(\mathcal{M}_{t_c,T})$ may not be convex in $\mathbb{R}^2$ for one or more $t$. 
To remedy this, one might compute an $\alpha$-complex\footnote{An $\alpha$-complex \cite{Edelsbrunner1983} is a subcomplex of a Delaunay triangulation often used to mesh non-convex domains. It requires a parameter value which restricts the triangular faces of the mesh to have radii no larger than the inverse of the parameter value.}, but this requires an \emph{a priori} non-obvious choice of the parameter $\alpha$ to control the non-convexity of the mesh.

We will instead use a parameter-free approach that utilizes the fact that we wish to mesh a subdomain $\Phi^{t-t_c}_{t_c}(\mathcal{M}_{t_c,T})$ of a larger domain $\Phi^{t}_{0}(\mathcal{M})$. 
We compute the Delaunay triangulation of 
$\Phi^{t}_{0}(\mathcal{M})$, and then select the mesh that contains $\{x_{i}^{t}\}_{i \in I_{t_c,T}}$ as vertices.
To create these meshes we express a point $x^{t}_{i} = (\alpha^{t}_{i}, \beta^{t}_{i}) \in \Phi^{t}_{0}(\mathcal{M})$ in three-dimensional Cartesian coordinates
\begin{equation}\label{spherical_polar}
y^{t}_{i} = \left( R_{\rm Earth} \cos \left(\alpha^{t}_{i}\right) \cos \left(\beta^{t}_{i} \right), R_{\rm Earth} \sin \left( \alpha^{t}_{i} \right) \cos \left( \beta^{t}_{i} \right), R_{\rm Earth} \sin \left(\beta^{t}_{i} \right) \right),
\end{equation}
where $R_{\rm Earth}=6,378$ kilometers. 
The pair $x_i^t$ and the triple $y_i^t$ represent the same element of $\Phi^t_0(\mathcal{M})$ in different coordinate systems.
Since our meshes lie near the surface of a sphere modeling the Earth, by meshing the three-dimensional Cartesian coordinates, the poles and the periodic boundary in the longitudinal direction require no special treatment.
The algorithm to compute a mesh for $\Phi^{t-t_c}_{t_c}(\mathcal{M}_{t_c,T})$, including determining its boundary, is described in Algorithm \ref{mesh_boundary}.\\
\begin{algorithm}[H]
\caption{Input - chosen time $t \in \overline{W}(t_c,T)$, particle set $\{x_{i}^{t}\}_{i=1,\dots,n} =: P$, indices $I_{t_c,T}$; Output - mesh $(V,E)$ and boundary edges $\partial E$ of $\Phi^{t-t_c}_{t_c}(\mathcal{M}_{t_c,T})$.} \label{mesh_boundary}
\begin{enumerate}
\item Map the particle set $P \subset \Phi^{t}_{0}(\mathcal{M})$ onto a sphere of radius $r =R_{\rm Earth}$ using (\ref{spherical_polar}) and append the origin $(0,0,0)$, creating a new particle set $P' = \{ y^{t}_{i} \in \mathbb{R}^{3}\}_{i = 1,\dots, n+1}$.
\item Compute a 3D convex hull triangulation of $P'$ and delete all triangular faces containing vertex $n+1$ (the origin).
The remaining set of faces is denoted $\mathcal{F}=\{\mathcal{F}(1),\ldots,\mathcal{F}(F)\}$, where for $f=1,\ldots,F$, $\mathcal{F}(f)\in\{1,\ldots,n\}^3$ is a triple of distinct vertex indices.
\item Let $\mathcal{F}'=\{\mathcal{F}(f)\in\mathcal{F} : \ \mathcal{F}(f)\subset I_{t_c,T}\}$ be the set of faces whose vertices all lie in $\Phi^{t-t_c}_{t_c}(\mathcal{M}_{t_c,T})$.
The indices of all vertices lying in $\Phi^{t-t_c}_{t_c}(\mathcal{M}_{t_c,T})$ is $V=\{i:\exists f, i\in \mathcal{F}(f)\subset\mathcal{F}'\}$.
\item By an edge we mean an unordered pair $(i_1,i_2)$, $i_1,i_2=1,\ldots,n$. 
The set of all edges connecting vertices in $\Phi^{t-t_c}_{t_c}(\mathcal{M}_{t_c,T})$ is $E=\{(i_1,i_2): \exists f, \{i_1,i_2\}\subset \mathcal{F}(f)\subset \mathcal{F}'\}$.
\item We define the boundary edges $\partial E:=\{(i_1,i_2): \exists! f, \{i_1,i_2\}\subset\mathcal{F}(f)\subset\mathcal{F}'\}$ as those edges belonging to a single face in $\mathcal{F}'$.
\item Output the mesh $(V,E)$, and the indices of boundary edges $\partial E$.
\end{enumerate}
\end{algorithm}

\begin{figure}[h]
\centering
\includegraphics[width=1\linewidth]{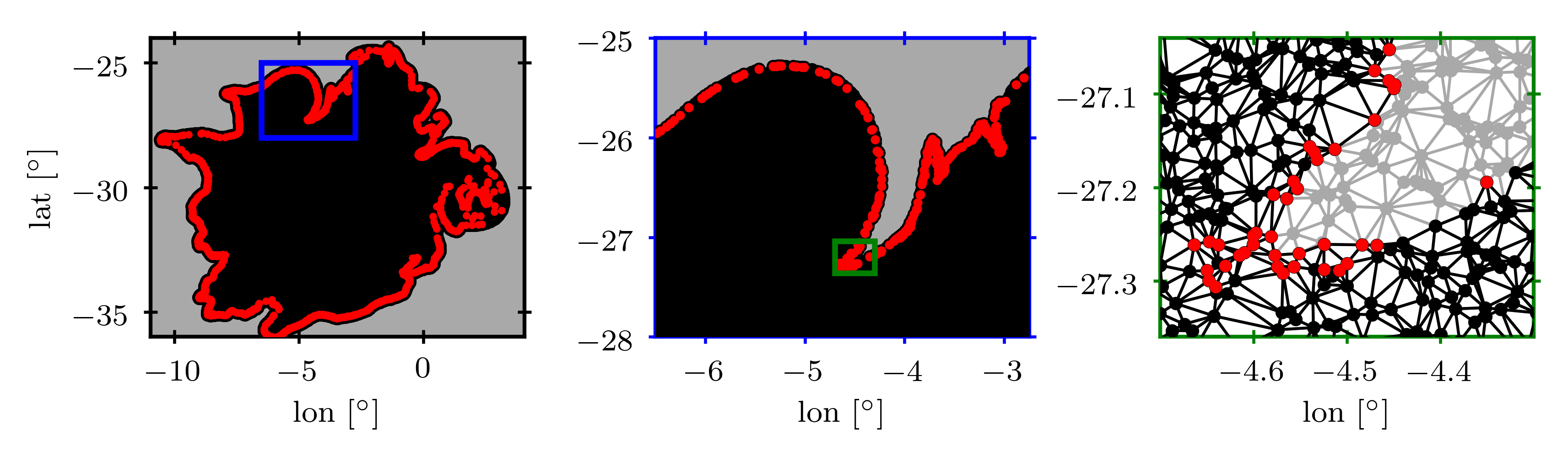}
\caption{The mesh of $\Phi^{-19}_{50}(\mathcal{M}_{50,100})$ is shown in black, computed as a sub-triangulation of $\Phi^{31}_{0}(\mathcal{M})$. The red particles are boundary particles, representing $\Phi^{-19}_{50}(\partial \mathcal{M}_{50,100})$, computed from the mesh of $\mathcal{M}_{50,100}$. A section of the mesh of $\Phi^{31}_{0}(\mathcal{M})$ is shown in gray and black.}
\label{fig:boundary_selection}
\end{figure}

To illustrate, a mesh of $\Phi^{-19}_{50}(\mathcal{M}_{50,100})$ is shown in Figure \ref{fig:boundary_selection}. The mesh of $\Phi^{-19}_{50}(\mathcal{M}_{50,100})$ is shown in black, and the red particles are the boundary particles computed from the mesh of $\mathcal{M}_{50,100}$, representing $\Phi^{-19}_{50}(\partial\mathcal{M}_{50,100})$. The mesh of $\Phi^{31}_{0}(\mathcal{M})$ is the combination of gray and black meshes.

\subsection{Numerical solution of the dynamic Laplacian eigenproblem} \label{sec:atom}
The numerical implementation of the dynamic Laplacian construction in section \ref{sec:DL} uses the adaptive transfer operator method developed in \cite{Froyland2018}. In this section we summarize this method and its implementation.
Within the time interval $W(t_c,T)$, we have a discrete set of times $t\in \overline{W}(t_c,T) := W(t_c,T)\cap\tau$.
In the description below, for each $t$ we reindex the points $\{y^t_i\}$, $i\in I_{t_c,T}$ to create the consecutively indexed points $\{\hat y^{t}_{k}\}$, $k=1,\ldots,N$, where  $N=|I_{t_c,T}|$.
The index $k$ is a reindexing of the global trajectory index $i$ that we will use to make the matrix constructions below clearer.
For each $t\in\overline{W}(t_c,T)$, let $V_{N}^{t} \subset H_{0}^{1}(\Phi^{t-t_c}_{t_c}(\mathcal{M}_{t_c,T}))$ be a finite-dimensional approximation space with basis $\{\varphi_{1}^{t}, \dots, \varphi_{N}^{t}\}$ of real-valued functions with domain $\Phi^{t-t_c}_{t_c}(\mathcal{M}_{t_c,T})$.
Define (very sparse) $N\times N$ matrices $D^t, M^t$, where the conditions are applied sequentially from first to last:
\begin{equation} \label{stiffness_matrix}
D_{kl}^{t} = \begin{cases}
					1 &\text{if }k=l,\\
					0 &\text{if }\hat y_k^{t_c+t} \in \partial\mathcal{M}_{t_c,T}\text{ or } \hat y_l^{t_c+t} \in \partial\mathcal{M}_{t_c,T},\\
					\int_{\Phi^{t-t_c}_{t_c}(\mathcal{M}_{t_c,T})} \nabla \varphi^{t}_k \bullet \nabla \varphi^{t}_l\ \text{d}l &\text{otherwise,}
					\end{cases}
\end{equation}
and,
\begin{equation} \label{mass_matrix}
M^{t}_{kl} = \begin{cases}
				0 &\text{if }\hat y_k^{t_c+t} \in \partial\mathcal{M}_{t_c,T}\text{ or }\hat y_l^{t_c+t} \in \partial\mathcal{M}_{t_c,T},\\
				\int_{\Phi^{t-t_c}_{t_c}(\mathcal{M}_{t_c,T})} \varphi_k^{t} \cdot \varphi_l^{t}\ \text{d}l &\text{otherwise.}
				\end{cases}
\end{equation}
Further define the $N\times N$ matrices
\begin{equation*}
D = \frac{1}{|\overline{W}({t_c,T})|} \underset{t \in \overline{W}({t_c,T})}{\sum}D^{t}, \text{ and }
M = \frac{1}{|\overline{W}({t_c,T})|} \underset{t \in \overline{W}({t_c,T})}{\sum}M^{t}.
\end{equation*}

The discrete eigenproblem corresponding to (\ref{weak_eigenproblem}), but replacing the interval $[-T/2,T/2]$ with the window $\overline{W}(t_c,T)$, is to find pairs $(\lambda, v), \lambda \in \mathbb{C}, v = \sum_{k=1}^{N} u_{k} \varphi_{k}^{t_c} \in V_{N}^{t_c}$, such that
\begin{equation}\label{discrete_eigenproblem}
-Du = \lambda M u,
\end{equation}
where $u = (u_1, \dots, u_N)^{\top} \in \mathbb{C}^{N}$.
For a fixed $t\in\overline{W}(t_c,T)$ the basis elements $\varphi_k^t$ will be standard piecewise affine ``hat'' functions, representing a nodal basis on a mesh of the trajectories $\hat y_{1}^{t}, \dots, \hat y_{N}^{t} \in \Phi^{t-t_c}_{t_c}(\mathcal{M}_{t_c,T})$ given by Algorithm \ref{mesh_boundary}; i.e.\ triangular $P_1$ Lagrange elements. 
Note that we create independent meshes for each $t\in\overline{W}(t_c,T)$;  at no time do we evolve a mesh with $\Phi^t_{t_c}$ or need to estimate $\Phi^t_{t_c}$ if we only have trajectory information.

The integrals in (\ref{stiffness_matrix}) and (\ref{mass_matrix}) are computed as two-dimensional integrals as follows.
Each triangular element of the mesh output by Algorithm \ref{mesh_boundary} is a triangle in $\mathbb{R}^3$.
Denoting the side lengths of this (positively oriented) triangle by $\ell_1,\ell_2,\ell_3$, we create a copy of this triangle in the plane, with vertices given in Cartesian coordinates by $(0,0), (\ell_1,0), (\ell_2\cos\theta,\ell_2\sin\theta)$, where $0<\theta<\pi$ is the angle made moving counterclockwise from the edge of length $\ell_1$ to the edge of length $\ell_2$.
The supports of $\varphi^t_k$ and $\varphi^t_l$ mostly do not intersect, and on the few occasions that they do, they intersect on the union of a small number of triangles, over which the contributions to the integrals in (\ref{stiffness_matrix}) and (\ref{mass_matrix}) can be evaluated analytically \cite{Froyland2018}.

We provide a compact Python implementation based on the code developed in \cite{Froyland2018} at \url{https://github.com/michaeldenes/FEMDLpy}. The modified algorithm is described in Algorithm \ref{atom_algorithm}.\\

\begin{algorithm}[h]
\caption{Input - time window $\overline{W}(t_c,T)$, indices $I_{t_c,T}$ of $\mathcal{M}_{t_c,T}$, trajectories $\{x_{i}^{t}\}_{t\in\mathcal{T}}, i\in I_{t_c,T}$; Output - Dominant eigenvalue and eigenvector $(\lambda, u)$ of $\Delta^{D}_{W(t_c,T)}$.} \label{atom_algorithm}
\begin{enumerate}
\item For each $t \in \overline{W}({t_c,T})$,
	\begin{enumerate}[label=(\alph*)]
	\item Apply Algorithm \ref{mesh_boundary} to obtain a mesh at time $t$.
	\item Compute $D^{t}$ using (\ref{stiffness_matrix}) and $M^{t}$ using (\ref{mass_matrix}) as described above.
	\end{enumerate}
\item Form $D$ and $M$, and solve the eigenproblem (\ref{discrete_eigenproblem}), for the leading eigenvalue $\lambda$ and corresponding eigenvector $u$.  Scale $u$ so that $\|u\|_\infty=1$.
\item Output $\lambda$, $u$.
\end{enumerate}
\end{algorithm}

\begin{figure}[h]
\centering
\includegraphics[width=1\linewidth]{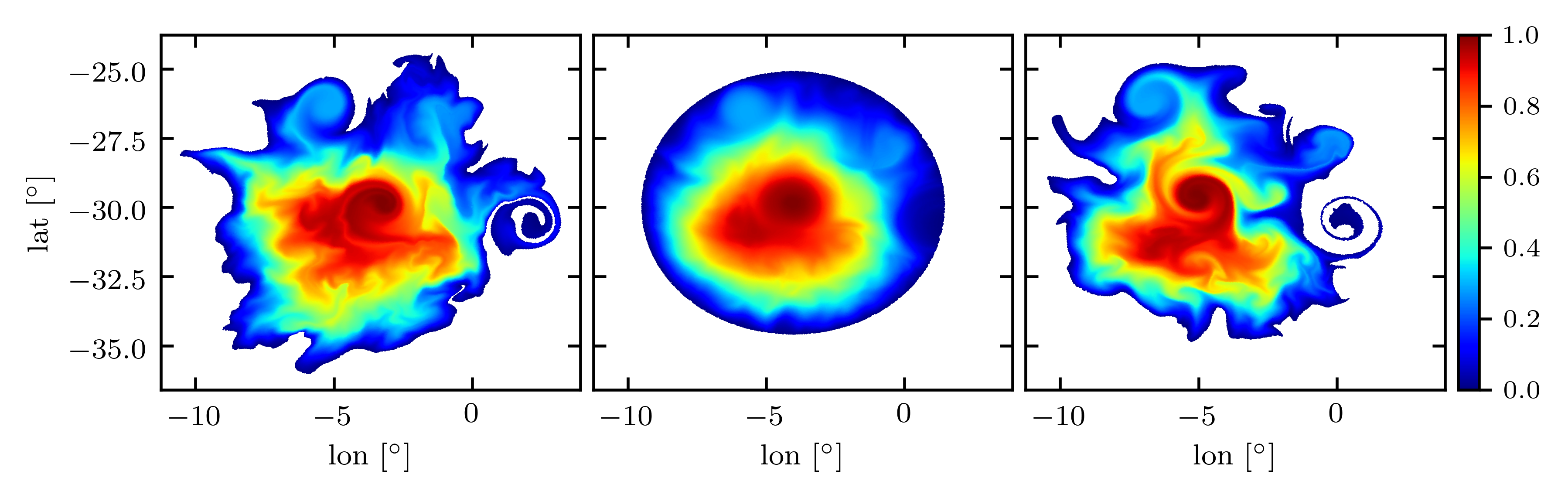}
\caption{(Center) The dominant eigenfunction $v$ of $\Delta^{D}_{W(50,38)}$ is plotted on a mesh of $\mathcal{M}_{50,100}$. (Left) The pull-back $(\Phi^{19}_{50})^{*}v$ is plotted on $\Phi^{-19}_{50}(\mathcal{M}_{50,100})$ and (Right) the push-forward $(\Phi^{19}_{50})_{*}v$ is plotted on $\Phi^{19}_{50} (\mathcal{M}_{50,100})$.}
\label{fig:eigenfunction_example}
\end{figure}

An example of an eigenfunction $v$ is shown in Figure \ref{fig:eigenfunction_example} (Center).
We always ensure that positive values of $v$ correspond to the core of the eddy by multiplying $v$ by $-1$ if necessary.

\subsection{Sequences of finite-time coherent sets from sliding windows}
\label{sec:3.5}
We now describe how to compute and track a sequence of finite-time coherent sets from a sequence of sliding windows. We start with a sequence of sliding windows $\{W(t_c,T)\}_{t_c \in \{t_{\min},\dots, t_{\max}\}}$, where the sequence of center times is spaced $1$ unit of time apart. As we slide the window from one center time $t_c$ to the next center time $t_c+1$, we wish to find a natural way to ``match'' the coherent set at $t_c$ to the one at $t_c+1$.
For the window $W(t_c,T)$ with center time $t_c$ and window length $T$, Algorithm \ref{mesh_boundary} provides a mesh of $\mathcal{M}_{t_c,T}$ and Algorithm \ref{atom_algorithm} provides the dominant normalized eigenvector $u$ and eigenfunction $v_{t_c}:\mathcal{M}_{t_c,T}\to\mathbb{R}$,  $v_{t_c} = \sum_{k=1}^{N} u_{k} \varphi_{k}^{t_c}$, corresponding to the leading eigenvalue $\lambda$ of the dynamic Laplacian $\Delta^{D}_{W(t_c,T)}$.
At $t=t_c$, we select an 
initial contour value $a(t_c)$, with which we define the finite-time coherent set $A_{a(t_c)}(t_c) = \{ x \in \mathcal{M}_{t_c,T} : v_{t_c}(x) \ge a(t_c)\}$.

We now define the closest-matching finite-time coherent set for the next window $W(t_c+1,T)$ as $A_{a(t_c+1)}(t_c+1)= \{ x \in \mathcal{M}_{t_c+1,T} : v_{t_c+1}(x) \ge a(t_c+1)\}$, where
\begin{equation} \label{contour_value_tracking}
    a(t_c+1) := \frac{\int_{\Phi^{1}_{t_c}(\partial A_{a(t_c)}(t_c))} v_{t_c+1}\ \ \text{d} l}{\ell_{d-1}(\Phi^{1}_{t_c}(\partial A_{a(t_c)}(t_c)))} = \frac{\int_{\partial A_{a(t_c)}(t_c)} (\Phi^{1}_{t_c+1})^{*} v_{t_c+1} \ \text{d}l}{\ell_{d-1}(\partial A_{a(t_c)}(t_c))}.
\end{equation}
The central term of (\ref{contour_value_tracking}) evolves forward the boundary of $A_{a(t_c)}(t_c)$ one time unit and averages the value of $v_{t_c+1}$ around this evolved boundary to define an updated contour value $a(t_c+1)$.
Equivalently, the term on the right in (\ref{contour_value_tracking}) computes the average value of the one-time-unit-pullback of $v_{t_c+1}$ around the boundary of $A_{a(t_c)}(t_c)$.
In our experiments, we begin at $t_c=t_{\min}$, calculate $v_{t_{\min}}$, and select an initial contour value $a(t_{\min})$.
We then iteratively apply (\ref{contour_value_tracking}):  i.e.\ (i) increment $t_{\min}$ to $t_{\min}+1$, (ii) calculate $v_{t_{\min}+1}$, and apply (\ref{contour_value_tracking}) to determine $a(t_{\min}+1)$, and repeating for $t_{\min}+2$, and so on. To carry out this averaging numerically we use the right-hand-side of (\ref{contour_value_tracking}).

We begin at $t_c=t_{\min}$ and select an initial contour value $a(t_{\min})$. We then compute the boundary of $A_{a(t_{\min})}(t_{\min})$ using the \texttt{tricontour} function in the Matplotlib package in python, which takes inputs of the eigenfunction $v_{t_{\min}}$ on the mesh of $\mathcal{M}_{t_{\min},T}$. This produces a list of points $z_{1}, \dots, z_{s}$ which lay along the boundary, from which we can compute the length of the boundary $\ell_{d-1}(A_{a(t_{\min})}(t_{\min}))$, ensuring we use great circle distances between points $z_{i}$ and $z_{i+1}$. We then linearly interpolate the pull-back of the eigenfunction $v_{t_{\min}+1}$ using the mesh $\Phi^{-1}_{t_{\min}+1}(\mathcal{M}_{t_{\min}+1,T})$ to the points $z_{1},\dots, z_{s}$, using the \texttt{LinearTriInterpolator} function. We compute the integral in the numerator of the right-hand-side of (\ref{contour_value_tracking}) using the trapezoidal rule. This determines $a(t_{\min}+1)$ which we use to determine $a(t_{\min}+2)$, and so on.

\subsection{Residence times of particles in a sequence of finite-time coherent sets} \label{sec:residence}

In order to quantify how well our sequence of finite-time coherent sets  $\{A_{a(t_c)}(t_c)\}_{t_c\in \{t_{\rm min},\dots, t_{\rm max}\}}$ transports fluid, we can compute residence times of the fluid in the sequence of sets.

For a selected center time $t_c$ the \emph{age} of a particle in the finite-time coherent set $A_{a(t_c)}(t_c)$ is the time the particle has been in the sequence of finite-time coherent sets up until (and including) time $t_c$, and the \emph{exit time} is the time from $t_c$ until the particle leaves the sequence. The \emph{residence time} of a particle is then defined as the total time the particle spends (without reentering) in the sequence of finite-time coherent sets, which is simply the sum of its age and exit times.

We now describe our residence time calculation for a fixed $t_c$.
Define the indices of particles in the finite-time coherent set $A_{a(t_c)}(t_c)$ as
\begin{equation}
    I_{t_c,T}^{a(t_c)} := \{ i \in I_{t_c,T}\ :\ x^{t_c}_{i} \in A_{a(t_c)}(t_c) \}.
\end{equation}
We may now compute the proportion of particles $p^-_{t_c}(t)$ that have remained for at least $t \ge 0$ days in the sequence of finite-time coherent sets prior to $t_c$, and the proportion of particles $p^+_{t_c}(t)$ that remain for at least $t \ge 0$ days in the sequence of finite-time coherent sets after $t_c$:
\begin{equation} \label{proportion_particles}
    p^-_{t_c}(t) := \frac{\left|\cap_{s=t_c-t}^{t_c} I_{s,T}^{a(s)}\right|}{\left|I_{t_c,T}^{a(t_c)}\right|} \qquad\mbox{ and }\qquad p^+_{t_c}(t) := \frac{\left|\cap_{s=t_c}^{t_c+t} I_{s,T}^{a(s)}\right|}{\left|I_{t_c,T}^{a(t_c)}\right|}.
\end{equation}
The mean age, exit, and residence time for fluid is given by 
\begin{equation}
    \label{meanageexit}
    \bar{\mathcal{A}}_{t_c}:=\sum_{t=0}^\infty p_{t_c}^-(t)t, \quad \bar{\mathcal{E}}_{t_c}:=\sum_{t=0}^\infty p_{t_c}^+(t)t,\quad\mbox{ and } \quad \bar{\mathcal{R}}_{t_c}:=\sum_{t=0}^\infty (p_{t_c}^-(t)+p_{t_c}^+(t))t.
\end{equation}

\section{Application to Agulhas Ring Dynamics} \label{sec:application}
In this section we will study the evolution and decay of an Agulhas ring. Agulhas rings are well-known examples of large ocean eddies which form due to instabilities in the Agulhas current and its retroflection \cite{Beal2011, DeRuijter1999}. They contribute to a significant portion of the Agulhas leakage, the transport of warm and salty water from the Indian ocean into the much cooler and fresher South Atlantic ocean \cite{Gordon1986, Biastoch2008}, which plays a key role in the maintenance of the global overturning circulation of the ocean \cite{Weijer2002, Biastoch2008a}.

Our study will focus on the Agulhas ring shown in Figure \ref{fig:intro_okubo_weiss}, with approximate initial center $[2^\circ \text{W},30.25^\circ \text{S}]$ at 30/12/2009, corresponding to $t_c=0$. We start by computing a representative maximal coherence timescale $\hat{T}$ for  $t_c=50$ in Section \ref{sec:telescoping}. We use this fixed maximal coherence timescale $\hat{T}$ to track the Agulhas ring as it evolves in Section \ref{sec:tracking}. To show our method is robust to the choice of window length $T$, we compare our results with tracking using
shorter ($T=10$) and longer ($T=60$) window lengths.
We also compute the residence times of surface water in our sequence of finite-time coherent sets for these window lengths and across three different superlevel sets of the corresponding eigenfunctions. Lastly, in Section \ref{sec:decay} we illustrate the dependence of the maximal coherence timescale on the window center time $t_c$.

\subsection{Experiment: A maximal coherence timescale of an Agulhas ring}\label{sec:telescoping}
To compute a maximal coherence timescale of the Agulhas ring at a specific time $t_c$ we construct a sequence of telescoping windows, centered on $t_c=50$ with window lengths equally spaced $2$ days apart; 
that is, $\{W(50,T)\}_{T\in\{10,12,\dots,100\}}$. To ensure the size and boundary of the domain of each window is the same, we use the domain $\mathcal{M}_{50,100}$ for all $T$, because $\mathcal{M}_{50,T}\subset \mathcal{M}_{50,100}$ for $T=10,12,\ldots,100$. The radius of this domain is approximately $520$km.
We solve the eigenproblem $\Delta^{D}_{W(50,T)}$ for each $T$. The dominant eigenvalue $\lambda$ and ratio $\lambda/T$ for each $T$ is shown in Figure \ref{fig:window_50_eigenvalues}, where one sees that the maximal coherence timescale for the Agulhas ring at $t_c=50$ is $\hat T=38$.

\begin{figure}[hbt]
\centering
\includegraphics[width=1\linewidth]{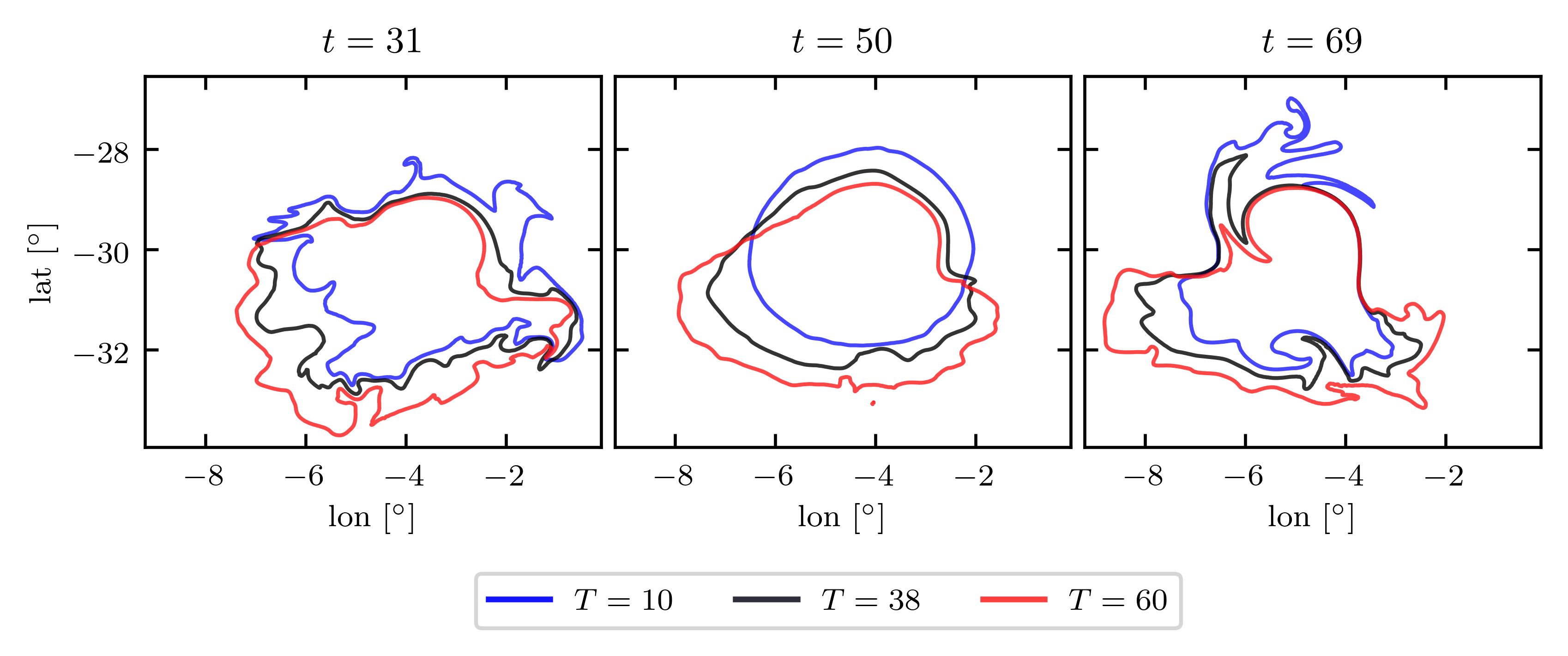}
\caption{The boundary $\partial A_{0.75}$ constructed from a level set of the dominant eigenfunction of $\Delta^{D}_{W(50,T)}$ (Center) for windows of lengths $T=10$ (blue), $T=38$ (black) and $T=50$ (red) plotted at $t=t_c=50$.
The pull-backs (resp.\ push-forwards) by 19 days are shown in the Left (resp.\ Right) images.}
\label{fig:eigenfunction_telescoping_comparison_contours}
\end{figure}

We plot the dominant eigenfunction of $\Delta^{D}_{W(50,38)}$ in Figure \ref{fig:eigenfunction_example}, along with a pull-back and push-forward by $\hat{T}/2 = 19$ days (the beginning and end of the window). 
In Figure \ref{fig:eigenfunction_telescoping_comparison_contours} (Center, black curve) we plot the level set of the eigenfunction in Figure \ref{fig:eigenfunction_example} at level set value $0.75$. Figure \ref{fig:eigenfunction_telescoping_comparison_contours} (Left, Right, black curves) show the backward (resp.\ forward) evolution of this level set under $\Phi$ for $\hat{T}/2=19$ days.
These two curves are equivalently characterized as the 0.75 level sets of the pull-back (resp.\ push-forward) of the eigenfunction, shown in Figure \ref{fig:eigenfunction_example} (Left, resp.\ Right).
As expected, the black curves in Figure \ref{fig:eigenfunction_telescoping_comparison_contours} remain short throughout the corresponding computation window $W(50,38)$.

We now compare this result with analogous calculations made with window lengths of $T=10$ and $T=60$.
The corresponding curves are plotted in Figure \ref{fig:eigenfunction_telescoping_comparison_contours}.
The black curve computed over the maximal coherence timescale $T=\hat{T}=38$ filaments significantly less than the blue curve computed over the shorter timescale $T=10$, and filaments an amount comparable to the red curve computed over the longer timescale $T=60$. 
Each of these results are expected because the blue curve computation over the shorter window makes no guarantee of coherence outside the interval $[t_c-10/2,t_c+10/2]$, while the red curve computation over the longer window $[t_c-60/2,t_c+60/2]$ does provide a guarantee for the pull-back/push-forward over 19 days.

\subsection{Experiment: Fixed-duration tracking of an Agulhas ring}\label{sec:tracking}

In order to quantify surface water escape from the Agulhas ring, we construct a sequence of sliding windows $\{W(t_c,38)\}_{t_c \in \{50,\dots,400\}}$, fixing $T=38$. Beyond $t_c=400$ the eddy is no longer recognizable by visual inspection in the vorticity-dominated region $Z(t_c)$. 

\subsubsection{Dependence of $\lambda$ and $\lambda(T)/T$ on center time $t_c$}
\begin{figure}[hbt]
\centering
\includegraphics[width=1\linewidth]{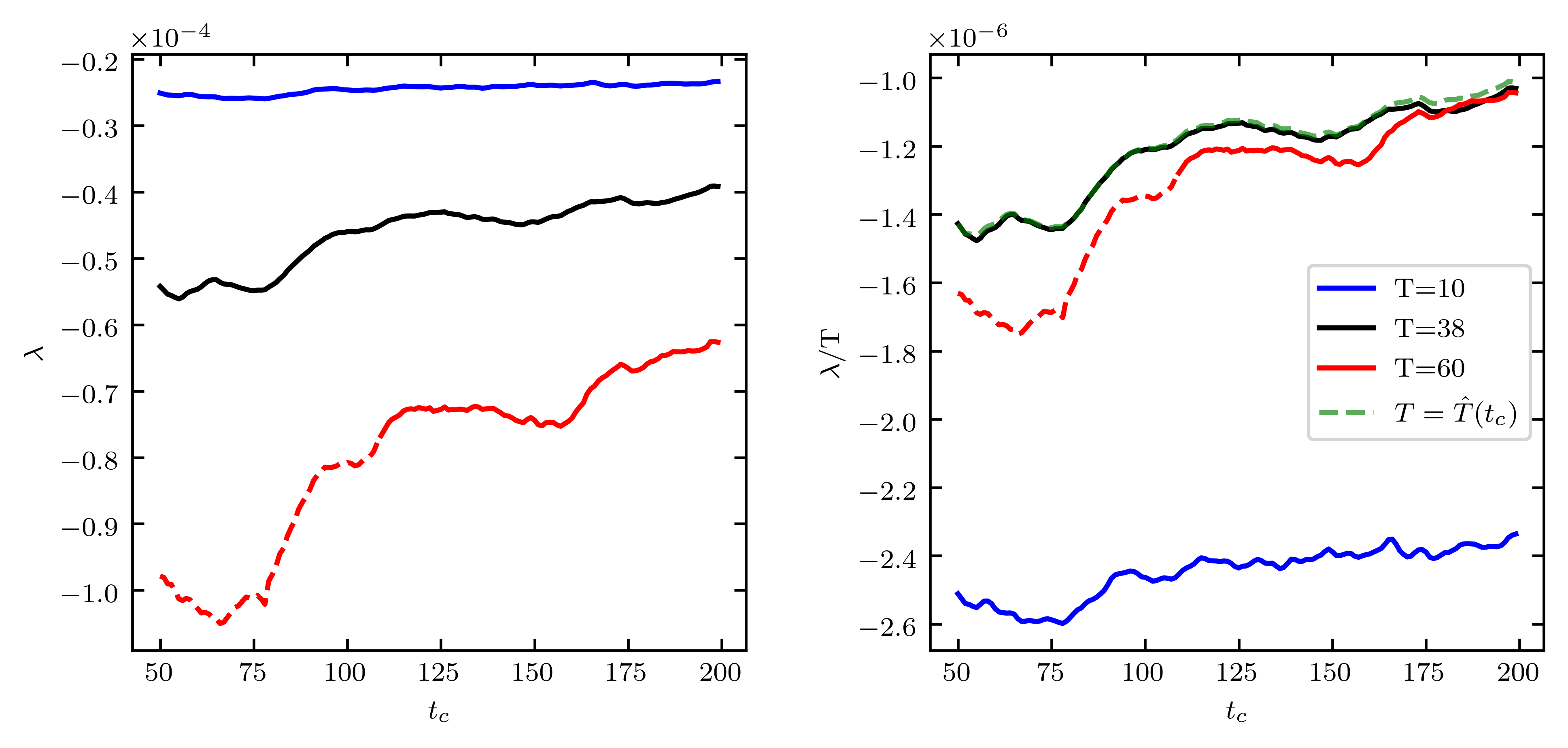}
\caption{The dominant eigenvalue $\lambda$ (Left), and mass loss rate $\lambda/T$ (Right) for three sequences of sliding windows with fixed window lengths $T=10$ (Blue), $T=38$ (Black), and $T=60$ (Red). The $T=60$ window is dashed for the initial center times where the eddy region does not appear as the most coherent region in the dominant eigenfunction. In the right panel, a fourth sequence of sliding windows with variable window length $T=\hat{T}(t_c)$ is shown in dashed green. This latter curve is by definition an upper bound for $\lambda(T)/T$ for all fixed choices of $T$.}
\label{fig:sliding_eigenvalues}
\end{figure}

We first plot the behavior of the leading eigenvalue $\lambda$ of $\Delta^D_{W(t_c,38)}$ as a function of $t_c$;  see Figure \ref{fig:sliding_eigenvalues} (Left, black).
The slight drift upwards indicates a slight strengthening of coherence over the time interval $[50,200]$, and that the maximal coherence timescale $\hat{T} = 38$ computed at $t_c=50$ corresponds to a slightly weaker phase of the eddy.
Analogous plots for $\lambda$ computed from $\Delta^D_{W(t_c,10)}$ and $\Delta^D_{W(t_c,60)}$ are shown in blue and red, respectively, also indicating a general strengthening of coherent dynamics as $t_c$ moves from 50 to 200.
In Figure \ref{fig:sliding_eigenvalues} (Right), we display the same information, but now scale $\lambda$ by $T=10, 38$, or $60$ as appropriate.
This normalization makes the curves directly comparable, and as discussed in Section \ref{sec:2.3} we interpret $\lambda/T$ as a \emph{rate} of mass loss per unit flow duration.
This figure shows a marked distinction between flow times $T=10$ and times $T=38,60$;  the former is much too short to correctly extract the coherent behavior of the eddy.
This calculation is consistent with the discussion concerning Figure \ref{fig:eigenfunction_telescoping_comparison_contours}.
In Figure \ref{fig:sliding_eigenvalues} (Right) we also plot the upper bound of $\lambda/T$ against $t_c$ (dashed green) by numerically computing $\hat{T}(t_c)$ for all $t_c\in[50,200]$, see Section \ref{sec:decay} for details. 
We see that $T=38$ represents a very close to optimal window length to minimize the rate of mass loss from our domain.

\subsubsection{Continuous dependence of the eigenvector $u_{t_c}$ of $\Delta_{W(t_c,38)}^D$ on $t_c$}
\begin{figure}[hbt]
\centering
\includegraphics[width=1\linewidth]{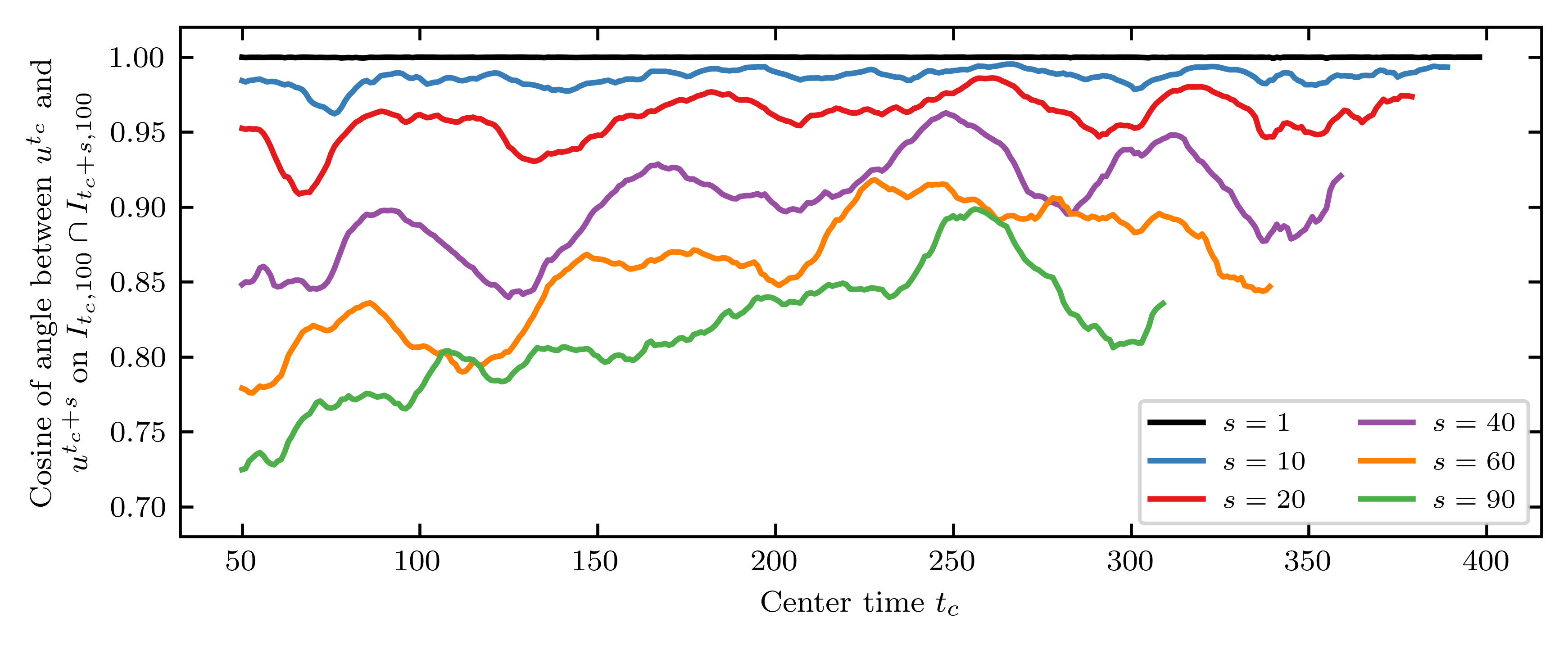}
\caption{The cosine of the angle between the dominant eigenvector $u^{t_c}$ of $\Delta^{D}_{W(t_c,38)}$ and the dominant eigenvector $u^{t_c+s}$ of $\Delta^{D}_{W(t_c+s,38)}$ on the intersection of their particle sets $I_{t_c,100} \cap I_{t_c+s,100}$, for a range of shift lengths $s \in \{1,10,20,40,60,90\}$.}
\label{fig:sliding_correlations_intersections}
\end{figure}

We demonstrate that our eigenvectors evolve in a continuous way as $t_c$ evolves. 
In the absence of domain changes, if the underlying time-dependent flow $\Phi^t$ is differentiable in space and time, then the eigenfunctions of $\Delta^D_{W(t_c,T)}$ will be continuously differentiable with respect to both $t_c$ and $T$ \cite{AFJ21}.
Our domain $\mathcal{M}_{t_c,100}$  also varies in a $C^1$ fashion with $t_c$ and therefore theoretically \cite{haddad2015} the leading eigenfunction should have $C^1$ dependence. 
To show that the (discretized) eigenvectors vary continuously, for the $T=38$ sequence we compute the cosine of the angle between successive eigenvectors on the region of their overlapping domains. This overlapping region, $\mathcal{M}_{t_c,100} \cap \Phi^{-1}_{t_c+1}(\mathcal{M}_{t_c+1,100})$, is approximated using the particle set $I_{t_c,100}\cap I_{t_c+1,100}$. Denote by $u^{t_c}$ the dominant eigenvector of $\Delta^{D}_{W(t_c,38)}$, of dimension $N = |I_{t_c,100}|$ whose elements are indexed by  $k=1,\dots,N$ which is a reindexing of the global trajectory index  $i \in I_{t_c,100}$, from the reindexing described in Section \ref{sec:atom}. Denote  the inner product restricted to indices in $\mathcal{I}\subset \{1,\ldots,N\}$ by $\langle \cdot,\cdot\rangle_{\mathcal{I}}$.
We calculate the cosine of the angle between $u^{t_c}$ and $u^{t_c+s}$ as
\begin{equation}
\frac{\langle u^{t_c},u^{t_c+s}\rangle_{\mathcal{I}}}{\sqrt{\langle u^{t_c}, u^{t_c}\rangle_{\mathcal{I}}} \sqrt{\langle u^{t_c+s}, u^{t_c+s} \rangle_{\mathcal{I}}}}, \ \text{where } \mathcal{I} = I_{t_c,100}\cap I_{t_c+s,100}.
\end{equation}
For $s=1$, this cosine is almost exactly 1 from $t_c=50$ to $t_c=399$, indicating an extremely smooth transition of eigenvectors from one day to the next. In other colors in Figure \ref{fig:sliding_correlations_intersections} we show analogous cosine values 
for $s=10, 20, 40, 60$, and $90$. 
These relatively high values also strongly indicate that the same persistent feature is being tracked.

\subsubsection{Dependence of water residence times on $t_c$ and eddy radius}
Using the formulae in Section \ref{sec:residence} we will quantify how well our tracked eddy $\{A_{a(t_c)}(t_c)\}_{t_c\in\{t_{\rm min},\ldots,t_{\rm max}\}}$ transports surface water by computing residence time statistics for water initialized at each $t_c$.
We will use three initial contour levels:  $a(t_c)=0.5, 0.75$, and $0.95$, to demonstrate how the eddy radius affects fluid transport.
Note that since our sequence of finite-time coherent sets begins at $t_{\min}=50$ and ends at $t_{\max}=400$, the largest possible directly calculable age is $t_c - 50$ and the largest possible directly calculable exit time is $400 -t_c$.
We therefore limit $t_c$ to the range $\{100,\ldots,350\}$ so that the directly calculable age and exit times are at least 50 days.

\begin{figure}
\centering
\includegraphics[width=1\linewidth]{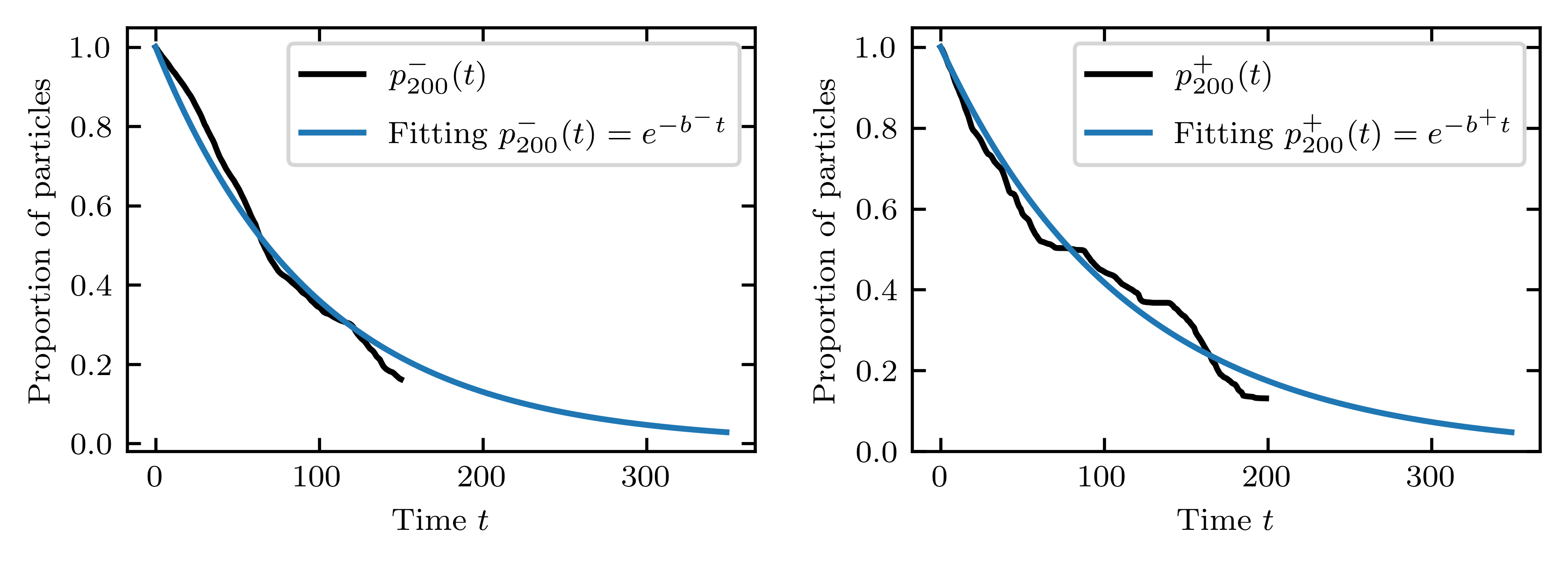}
\caption{A representative fit of the age (Left) and exit times (Right) at $t_c=200$ for initial contour value $a(50)=0.5$. Here we computed $b^{-} = 1.019 \times 10^{-2}$, and $b^{+} = 8.729\times 10^{-3}$.}
\label{fig:fitted_exponentials}
\end{figure}

To provide consistent estimates for age and exit times across different $t_c$, we will extrapolate age and exit probabilities with an exponential\footnote{The choice of exponential decay is a natural and convenient one, relevant for Markov processes (see e.g.\ residence time calculations for Antarctic gyres as almost-invariant sets in a Markov model of the Southern ocean \cite{dellnitz09}) or uniformly hyperbolic chaotic dynamics (see e.g.\ \cite{DemersYoung06}).  
Another possibly relevant choice  would be algebraic (power law) decay, but \emph{ad hoc} modifications are necessary to avoid a singularity at $t=0$. The exponential ansatz is more conservative in terms of expected residence times because of its thinner tail.} decay.
More precisely, using data at $t\in\{0,1,\ldots,t_c-t_{\rm min}\}$ we perform a nonlinear least-squares fit for $p^-_{t_c}(t) = e^{-(b^-)t}$, $b^->0$, where $p^-_{t_c}(t)$ is directly calculated in (\ref{proportion_particles}).  Similarly, using data at $t\in\{0,1,\ldots,t_{\rm max}-t_{c}\}$ we fit  $p^+_{t_c}(t) = e^{-(b^+)t}$, $b^+>0$.
We have suppressed the $t_c$-dependence of $b^-,b^+$ for brevity.
Figure \ref{fig:fitted_exponentials} shows a representative exponential fit of the age and exit times at $t_c=200$. Note that according to this fit, the ``half-life'' of the age (resp.\ exit time) of water in the tracked sequence of eddies is around 68 days (resp.\ 79 days).

Interpreting $p^-_{t_c}(t)$ (resp.\ $p^+_{t_c}(t)$) as the probability density of an exponentially distributed random variable $X_{t_c}^-$ (resp.\ $X_{t_c}^+$), assuming that $X_{t_c}^-$ and $X_{t_c}^+$ are independent with distinct means, one has that $X_{t_c}:=X_{t_c}^-+X_{t_c}^+$ is distributed
hypoexponentially: $X_{t_c} \sim \text{Hypo}(b^-,b^+)$, with mean $\bar{X}_{t_c} = 1/b^- + 1/b^+$.
The cumulative distribution function of $X_{t_c}$, which will be useful for calculating percentiles shortly, is given by
\begin{equation}
\label{hypocum}
    F_{t_c}(t) =  1 -\frac{b^+}{b^+-b^-}e^{-b^-t} + \frac{b^-}{b^+-b^-}e^{-b^+t},\quad t\ge 0.
\end{equation}

\begin{figure}
\centering
\includegraphics[width=1\linewidth]{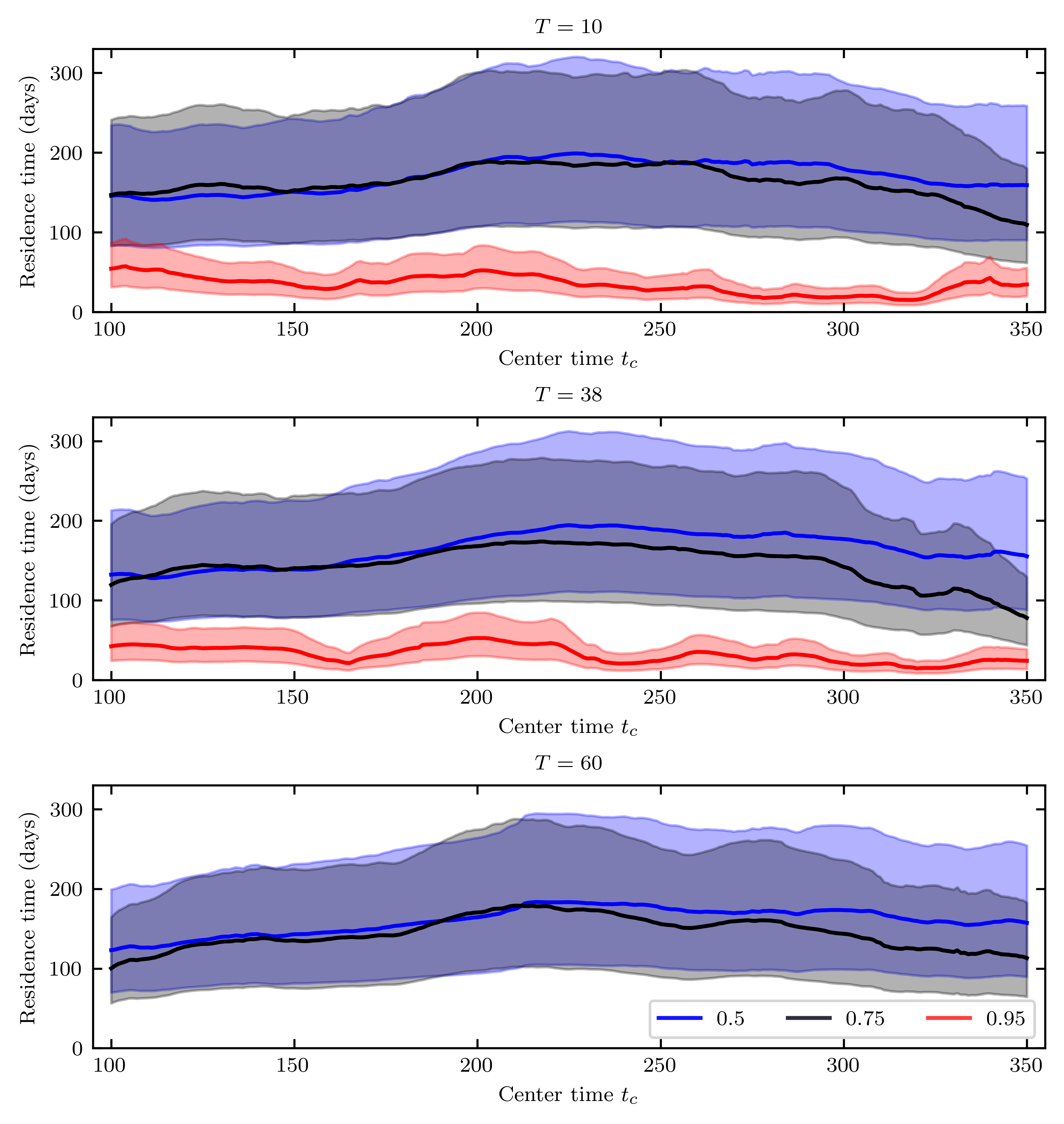}
\caption{Residence times for particles contained in three sequences of finite-time coherent sets, with initial superlevel set values at $t_c=50$ given by $a(50) = 0.5$ (Blue), $a(50) = 0.75$ (Black), $a(50) = 0.95$ (Red).  
The residence time calculation is repeated for window lengths $T=10$ (Upper panel), $T=38$ (Center panel) and $T=60$ (Lower panel). The solid line represents median residence time, and the shaded region represents the 25\% to 75\% percentiles. The $0.95$ contour of the $T=60$ window is omitted as the eddy region does not appear as the most coherent region in the dominant eigenfunction.}
\label{fig:residence_combined}
\end{figure}

Figure \ref{fig:residence_combined} shows the variation of the $25^{\rm th}$-percentile, median, and $75^{\rm th}$-percentile of residence times with $t_c$, calculated from (\ref{hypocum}), for initial contour values $a(t_c)=0.5, 0.75$, and $0.95$, and for window lengths $T=10, 38,$ and $60$.

For the maximal coherence timescale $T=38$ (center panel), the particles contained in the $0.5$ contour (blue) begin with a median residence time of $132$ days, 
increase to a maximum median residence time of $194$ days at $t_c=225$, and remain relatively stable after that, slightly decreasing over time, consistent with a slowly decaying eddy.
The particles contained in the $0.75$ contour (black) follow a similar pattern:  once the maximum median residence time of $174$ days at $t_c=218$ is reached, the median residence time begins to fall, again consistent with a decaying eddy. 
The $0.95$ contour (red) represents a region similar to the inner core of an eddy. 
The median residence time of particles contained in the $0.95$ contour is much smaller than the other two contours.
This suggests that in this eddy, no long-lived material inner core exists, and that what may be deemed as the ``inner core'' mixes with the quasi-coherent outer ring of the eddy itself. 
If a long-lived material inner core did exist, we would expect significantly longer median residence times.
Similar observations can be made with both the shorter window length $T=10$, and the longer window length $T=60$.

We have omitted the $0.95$ contour from the $T=60$ window length as the eddy region does not appear as the most coherent region in the dominant eigenfunction. 
This is not surprising, given that the median residence times for the 0.95 contour for $T=10$ and $T=38$ never exceed 60 days;  the 60-day window is likely simply too long for persistent coherent dynamics near the eddy core.
This is consistent with our finding that no long-lived material inner core exists for this particular eddy.

\subsection{Experiment: Variation of  maximal coherence timescale over time}\label{sec:decay}
\begin{figure}[hbt]
\centering
\includegraphics[width=1\linewidth]{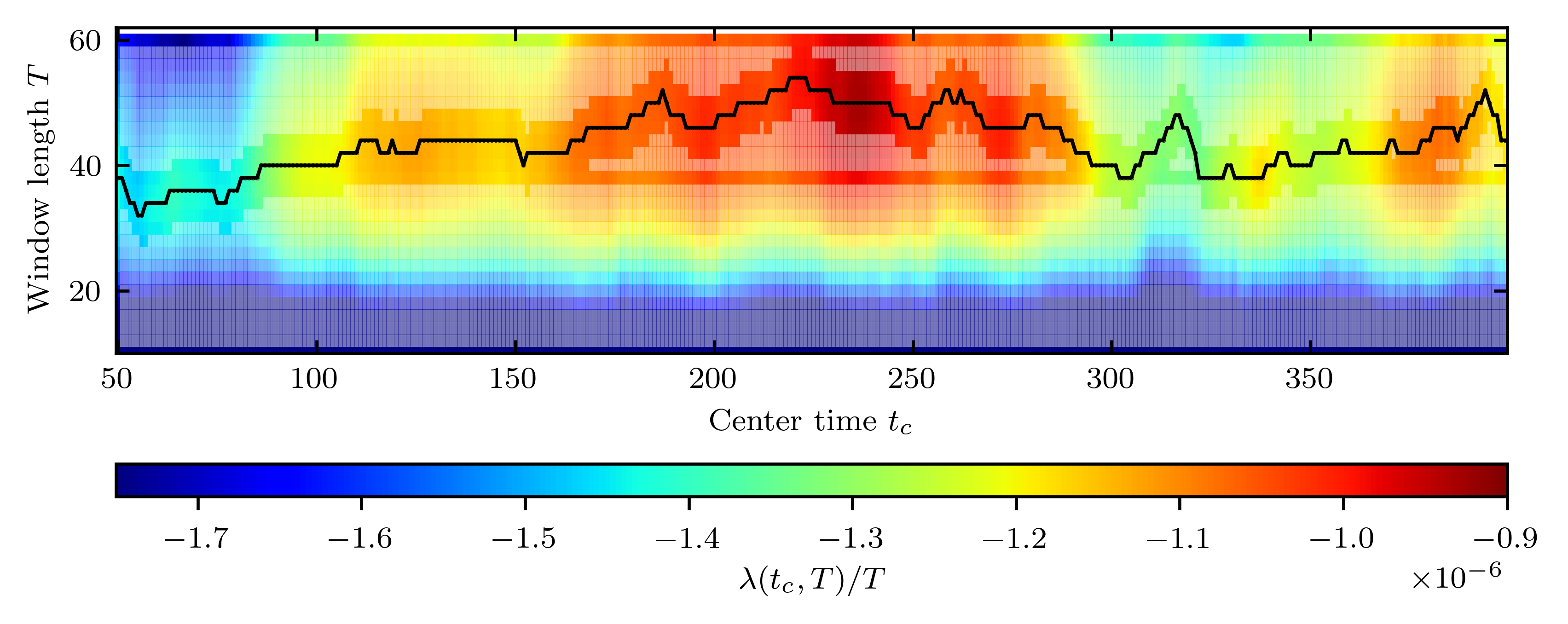}
\caption{The black line shows the maximal coherence timescales $\hat{T}(t_c)={\rm argmax}_{T}\lambda(t_c,T)/T$ as a function of $t_c$. The ratio $\lambda(t_c,T)/T$ is plotted for all center times $t_c \in \{50, 51, \dots,400\}$ and window lengths $T \in \{10, 12, \dots, 60\}$. Values of $\lambda(t_c,T)/T$ were explicitly computed at (i) all $T\in\{10,12,\dots,60\}$ at $t_c=50$, (ii) at $T=10,38,$ and $60$ for all $t_c \in \{50, 51, \dots, 400\}$, and (iii) in a radius of four days centered at the previous $\hat{T}$ for each $t_c$. A cubic spline interpolation across $t_c$ (shown in lighter color) was used to estimate values at combinations of $(t_c,T)$ not explicitly computed. To highlight the variation of $\lambda(t_c,T)/T$ we have truncated the colorbar at $-1.75\times 10^{-6}$.}
\label{fig:optimal_tracking_lambda}
\end{figure}

In this section we illustrate how the maximal coherence timescale $\hat{T}$ varies with $t_c$.
At $t_c=50$ we have already computed $\hat{T}=38$.
Because we expect smooth variation \cite{AFJ21} of $\lambda(t_c,T)$ with respect to both $t_c$ and $T$, as we advance $t_c$ to $t_c+1$ we need only compute $\lambda(t_c+1,T)$ for $T$ in a small range of window lengths centered around the previous $\hat{T}$.
We therefore limited our search to a window of radius four days centered at the previous $\hat{T}$.
Figure \ref{fig:optimal_tracking_lambda} plots $\hat{T}(t_c)$ vs. $t_c$ (black line). 
The black line is displaying the same information as the green dashed line in Figure \ref{fig:sliding_eigenvalues} (Right), but with a $t_c$ axis extended to $t_c=400$.
Values of $\lambda(t_c,{T})/{T}$ for combinations of $t_c$ and $T$ are shown in color.

Following the general upward drift in maximal coherence timescale $\hat{T}$ --- corresponding to the eddy strengthening  already noted in Figure \ref{fig:sliding_eigenvalues} (Right) --- we see a downward trend in $\hat{T}$ from $t_c=235$ to $t_c=370$, corresponding to eddy weakening.
We emphasize that the quantities $\hat{T}$ and $\lambda(t_c,\hat{T})/\hat{T}$ are independent.
It is possible to have relatively short coherence timescale $\hat{T}$ with strong coherence \emph{over that timescale} or to have a relatively long coherence timescale $\hat{T}$ with weak coherence.
Indeed, at around $t_c=315$ we see that $\hat{T}$ lengthens, while $\lambda(t_c,\hat{T}(t_c))/\hat{T}(t_c)$ experiences a slight drop, corresponding to weaker coherence.

\section{Discussion and Conclusion} \label{sec:discussion}

Accurate estimation of fluid trapping, leakage, and entrainment by ocean eddies is necessary to quantify their role in transporting mass, heat, and biogeochemical tracers and hence, their impact on the climate system. Estimates of transport by long-lived material inner cores  provide a lower bound on overall eddy transport \cite{Haller2013,Wang2015,Wang2016,Haller2016,Abernathey2018}. However, few studies \cite{Early2011,Froyland2015c} have estimated material transport by the outer ring, or by an eddy with no distinguishable long-lived inner core.

We propose an objective Lagrangian method to quantify the transport statistics of surface water by both the inner core and the outer ring of an ocean eddy. Our method is based on a sliding sequence of windows from which we track a sequence of finite-time coherent sets, namely, superlevel sets of the dominant eigenfunction of the dynamic Laplace operator \cite{Froyland2015a}. A systematic study of the effect of the window length (or flow duration) on the dominant eigenvalue allows us to define the maximal coherence timescale $\hat{T}$ as the window length that minimizes the rate of mass loss per unit flow time. We extend the numerical approach developed by \cite{Froyland2018} to construct meshes of (non-convex) domains on the surface of a sphere and describe a new approach to track finite-time coherent sets across overlapping windows of length $\hat{T}$.

The resulting sequence of finite-time coherent sets is capable of capturing quasi-coherent features --- including both the inner core and outer ring of an eddy --- that persist for times much longer than their material coherence timescale. Thus, the method is able to track an ocean eddy throughout its lifetime while explicitly allowing for fluid to be entrained by, and leak from, the eddy.

In our study we considered a region of the South Atlantic, where Agulhas rings regularly form, evolve, and decay. Focusing on a single eddy feature at a particular timestamp, we found a representative maximal coherence timescale of $\hat{T}=38$ days.
Using this timescale, we estimated transport, leakage, and entrainment by the eddy. By computing the dominant eigenfunctions of dynamic Laplace operators on a sequence of sliding windows of length 38 days, we found that the eddy-like feature identified by the initial dominant eigenfunction persisted over at least 300 days.
For the initial window, we defined three finite-time coherent sets, that represented a potential inner core and two outer rings. We tracked sequences of these finite-time coherent sets and estimated the distributions of residence times of particles contained within.
Although strict material transport was only imposed for 38 days, the sequence of identified features recorded median residence times of 128--194 days.
We made analogous calculations for window lengths $T=10$ and $T=60$ for comparison.

Surprisingly, we found that the eddy exhibited no long-lived material inner core. This is evidenced by the short median residence times of the tracked inner core sequence for the $T=38$ window, ranging between $15$ and $52$ days. Additionally, the dominant eigenfunction of the initial $T=60$ windows were unable to capture the eddy as the most coherent region. In contrast, the much longer median residence times of the tracked outer ring sequences for all three window lengths suggested that, for this particular eddy, the outer ring was a significant contributor to the entrainment and retention of fluid by the eddy, while the inner core played no significant role in material transport.
The qualitative similarity between the median residence times of the two tracked outer ring sequences suggested that these outer rings are robust to the choice of window length, and that our method can capture the multi-timescale nature of this particular ocean eddy.

While the eddy in our study exhibited no coherent inner core, our results support recent findings \cite{Wang2015, Wang2016, Abernathey2018} that point to the potentially important contribution to eddy transport due to the quasi-coherent outer ring of eddies. 
Whether the lack of a coherent inner core is unusual, or is in fact typical of most eddies, is a  subject for further study. Calculations of residence times of a larger number of eddies, using the methods developed in this paper, may shed light on the significance of fluid entrainment and retention within the outer ring of eddies. This too is a subject for future work.

\section*{Acknowledgments}
The authors would like to thank Jason Atnip for useful suggestions regarding exponential decay of residence times, Christopher Rock for helpful discussions on the dynamic Buser inequality, and to the COSIMA consortium (www.cosima.org.au) for making available the ACCESS-OM2 suite of models (available at \url{https://github.com/COSIMA/access-om2}.)
This research was undertaken with the assistance of resources and services from the National Computational Infrastructure (NCI), which is supported by the Australian Government. MCD is supported by an Australian Government Research Training Program Scholarship, GF is partially supported by an ARC Discovery Project, and SRK acknowledges the support of the Australian Research Council (ARC) through grants LP170100498 and DP210102745 and the ARC Centre of Excellence for Climate Extremes (CLEX; CE170100023).

\section*{Conflict of Interest}
The authors declare that they have no conflict of interest.

\section*{Data Availability Statement}
The data and relevant code that support the findings of this study are available from the corresponding author upon reasonable request. The ACCESS-OM2 model output is hosted in the COSIMA Model Output Collection: doi:10.4225/41/5a2dc8543105a. The Parcels python framework is available at \url{https://github.com/OceanParcels/parcels}.

\appendix

\section{Further discussion of the maximal coherence timescale}
\label{sec:timescale}

To further motivate the formula (\ref{optimal_t}) for $\hat{T}$ we return to the classical transfer operator constructions for coherent set detection.
Reference \cite{Froyland2013} considers a transfer operator $\mathcal{P}_\epsilon^t$, which, in the notation of the present paper is constructed as $\mathcal{P}_\epsilon^tf=\mathcal{D}_\epsilon(\Phi_*^t f)$, where $\mathcal{D}_\epsilon f$ is an averaging operator representing the addition of small diffusion.
Alternatively, if $\Phi^t$ is the flow map of a vector field $\mathcal{V}$ one may consider the SDE $dx=\mathcal{V}(x,t)dt + \epsilon db$, where $b$ is standard Brownian motion.
The corresponding transfer operator (also denoted here by $\mathcal{P}_\epsilon^t$) has small diffusion added continuously in time.
In both the discrete and continuous applications of diffusion (see \cite{DJM16} for details on the latter), following \cite{Froyland2013}, one considers singular vectors of $\mathcal{P}_\epsilon^t$;  in other words, eigenfunctions of $(\mathcal{P}_\epsilon^t)^*\mathcal{P}_\epsilon^t$.

For a flow duration $T$, let $\Lambda_\epsilon(T)$ denote the largest nontrivial singular value\footnote{All such eigenvalues are real and lie in $[0,1]$ \cite{Froyland2013}.} 
of $\mathcal{P}_\epsilon^t$. 
With Neumann boundary conditions, total mass is conserved and the largest nontrivial singular value $\Lambda_\epsilon(T)<1$ is the second singular value.
The singular value $\Lambda_\epsilon(T)$ quantifies the amount of $L^2$ mixing that occurs over the finite time duration $T$ \cite{Froyland2013}.
This mixing is connected with the dominant coherent set over the time duration $T$ because mass inside this coherent set is prevented from mixing with mass outside the set. 
With Dirichlet boundary conditions, mass leaks through the boundary and the largest nontrivial singular value $\Lambda_\epsilon(T)<1$ is the leading singular value.
This eigenvalue may be identified with the $L^2$ fraction of mass (distributed like the leading (left) singular function) that remains in the domain after the finite time duration $T$.
Similar to the Neumann case, this leaking is connected with the dominant coherent set over the time duration $T$ because mass inside this coherent set is prevented from leaking out of the set and subsequently through the boundary of the domain. 

In each boundary condition setting, the closer $\Lambda_\epsilon(T)$ is to one, the less mixing or mass loss experienced, respectively. 
Therefore, if we wish to find a $T$ so that \emph{the mixing (resp.\ mass loss) per unit flow duration is least}, we should choose $T$ to
\begin{equation}
    \label{maxLam}
\mbox{maximize } \Lambda_\epsilon(T)^{1/T}.
\end{equation}

We now link these considerations to the dynamic Laplacian approach.
By Theorem 5.1 \cite{Froyland2015a}, for smooth $f$ one has:
\begin{equation}
    \label{oplimit}\lim_{\epsilon\to 0}\frac{(\mathcal{P}_\epsilon^t)^*\mathcal{P}_\epsilon^tf-f}{\epsilon^2}=\Delta^D_{W(0,T)}f.
\end{equation} 
Setting $f$ to be the second eigenfunction for $(\mathcal{P}_\epsilon^t)^*\mathcal{P}_\epsilon^t$, we expect
\begin{equation}
    \label{evallimit}
\lim_{\epsilon\to 0}\frac{\Lambda_\epsilon({T})^2-1}{\epsilon^2}=\lambda(T),
\end{equation}
where $\lambda(T)<0$ is the leading nontrivial eigenvalue of $\Delta^D_{W(0,T)}$.
Equation (\ref{oplimit}) is discussed for multiple time steps in \cite{banischkoltai} and the limit (\ref{evallimit}) is proved in a spectral (rather than pointwise) sense in \cite{KS21}, so that $\lambda(T)$ is indeed the leading nontrivial eigenvalue of $\Delta^D_{W(0,T)}$.
Rearranging (\ref{evallimit}) one obtains
$\Lambda_\epsilon(T)^2\approx 1+\epsilon^2\lambda(T)$ for small $\epsilon$ and taking logs, by Taylor we have
\begin{equation}
    \label{convert}
\log\Lambda_\epsilon(T)^2\approx \log(1+\epsilon^2\lambda(T))\approx \epsilon^2\lambda(T),
\end{equation}
for small $\epsilon$.
Using (\ref{convert}), we may re-express the (logarithm of the) per-unit-flow-duration rate from (\ref{maxLam}): $$\log\Lambda_\epsilon(T)^{1/T}=\frac{\log\Lambda_\epsilon(T)}{T}\approx 
 \frac{\epsilon^2}{2}\cdot\frac{\lambda(T)}{T}.$$
Thus, for small fixed $\epsilon$, \emph{maximizing $\Lambda_\epsilon(T)^{1/T}$ over $T$ corresponds to maximizing $\lambda(T)/T$ over $T$.}

\bibliography{paper}
\bibliographystyle{siam}
\end{document}